\documentclass[12pt,1p]{elsarticle}

\usepackage{lineno,hyperref,url}
\usepackage{graphicx,float}
\usepackage{makecell}
\usepackage{rotating,booktabs}
\usepackage{eqnarray,amsmath}
\usepackage{subcaption}
\usepackage{graphicx}
\modulolinenumbers[5]

\journal{Journal of Quantitative Spectroscopy and Radiative Transfer }
\begin{document}

\begin{frontmatter}

\title{Skyglow Changes Over Tucson, Arizona, Resulting From A Municipal LED Street Lighting Conversion}

\author[ida,cdss]{John C.~Barentine\corref{cor1}} 
\ead{john@darksky.org}

\author[noao,ida]{Constance E.~Walker}
\ead{cwalker@noao.edu}

\author[ica,comenius]{Miroslav Kocifaj}
\ead{Miroslav.Kocifaj@savba.sk}

\author[comenius]{Franti\v{s}ek Kundracik}
\ead{frantisek.kundracik@fmph.uniba.sk}

\author[uofa]{Amy Juan}
\ead{protecthimdag@gmail.com}

\author[sacto]{John Kanemoto}
\ead{kanemotojohn@hotmail.com}

\author[monrad]{Christian K.~Monrad}
\ead{chrismonrad@monradengineeringinc.com}

 \cortext[cor1]{Corresponding author}

 \address[ida]{International Dark-Sky Association, 3223 N.~1st Ave, Tucson, AZ 85710 USA}
 \address[cdss]{Consortium for Dark Sky Studies, University of Utah, Sterling Sill Center, 195 Central Campus Dr, Salt Lake City, UT 84112 USA}
 \address[noao]{National Optical Astronomy Observatory, 950 N. Cherry Ave, Tucson, AZ 85719 USA}
 \address[ica]{ICA, Slovak Academy of Sciences, D\'{u}bravsk\'{a} Road 9, 845 03 Bratislava, Slovakia}
 \address[comenius]{Faculty of Mathematics, Physics, and Informatics, Comenius University, Mlynsk\'{a} Dolina, 842 48 Bratislava, Slovakia}
 \address[uofa]{University of Arizona, Tucson, AZ 85719 USA}
 \address[sacto]{Natomas Unified School District, 1901 Arena Blvd., Sacramento, CA 95834 USA}
  \address[monrad]{Monrad Engineering, Inc., 1926 E. Fort Lowell Road Suite 200, Tucson, AZ 85719 USA} 

%
%
 \begin{abstract}
The transition from earlier lighting technologies to white light-emitting diodes (LEDs) is a significant change in the use of artificial light at night. LEDs emit considerably more short-wavelength light into the environment than earlier technologies on a per-lumen basis. Radiative transfer models predict increased skyglow over cities transitioning to LED unless the total lumen output of new lighting systems is reduced. The City of Tucson, Arizona (U.S.), recently converted its municipal street lighting system from a mixture of fully shielded high- and low-pressure sodium (HPS/LPS) luminaires to fully shielded 3000 K white LED luminaires. The lighting design intended to minimize increases to skyglow in order to protect the sites of nearby astronomical observatories without compromising public safety. This involved the migration of over 445 million fully shielded HPS/LPS lumens to roughly 142 million fully shielded 3000 K white LED lumens and an expected concomitant reduction in the amount of visual skyglow over Tucson. SkyGlow Simulator models predict skyglow decreases on the order of 10-20\% depending on whether fully shielded or partly shielded lights are in use. We tested this prediction using visual night sky brightness estimates and luminance-calibrated, panchromatic all-sky imagery at 15 locations in and near the city. Data were obtained in 2014, before the LED conversion began, and in mid-2017 after approximately 95\% of $\sim$18,000 luminaires was converted. Skyglow differed marginally, and in all cases with valid data changed by $<{\pm}$20\%. Over the same period, the city's upward-directed optical radiance detected from Earth orbit decreased by approximately 7\%. While these results are not conclusive, they suggest that LED conversions paired with dimming can reduce skyglow over cities.
 \end{abstract}
 


\end{frontmatter}

%
%
\section{Introduction}

The conversion of the world's lighting from conventional to solid-state lighting (SSL) technologies is among the most significant changes to the way we light our world at night since the invention of electric light itself. Environmental pollution from artificial light at night (ALAN) has already reached significant levels in many parts of the world.~\citep{Falchi2016} Improved luminous efficacy among SSL products is hypothesized to lead to a ``rebound'' effect, furthering global dependence on ALAN as cost savings are redirected into the deployment of additional outdoor lighting.~\citep{Tsao2010,SaundersTsao2012,TsaoWaide2013,Borenstein2014} The conversion to SSL has also brought significant changes to the spectrum of artificial light radiated into the global nighttime environment, shifting a considerable amount of emission to shorter wavelengths. A number of known and suspected hazards to wildlife ecology and human health have been identified that are thought to result from exposure to short-wavelength ALAN.~\citep{Gaston2015,Zubidat2017,Lunn2017}

The spectrum shift in new SSL systems is also thought to yield increases to skyglow, which is the diffuse luminescence of the night sky attributable to light emitted from sources on the ground that is scattered back toward the ground from molecules and aerosols in the Earth's atmosphere. Enhanced short-wavelength light emissions associated with blue-rich white LED are subject to strong Rayleigh scattering in the atmosphere, resulting in higher scattering probabilities, associated with the formation of skyglow, than light of longer wavelengths. Radiative transfer models therefore predict that conversion from older technologies to solid-state lighting should result in more skyglow, even when the system outputs are matched lumen-for-lumen.~\citep{Luginbuhl2014} Clouds, fog, and other sources of opacity at optical wavelengths in the lower atmosphere amplify skyglow \citep{Kyba2011}, leading to higher sky luminance values and resulting increases in ambient light level at ground level in cities. As the world increasingly adopts SSL, we expect the associated problems to be exacerbated unless the lighting conversions involve corresponding reductions in the overall light levels employed. However, relatively few communities to date have experimented with reducing lighting levels as they convert their municipal lighting systems to solid state.

It is often held in media coverage of SSL conversions that moving from legacy lighting technologies to light emitting diode (LED) lighting will reduce ``light pollution'' because almost all new luminaire designs are fully shielded and LEDs are highly directional light sources.~\citep{ThomsonAnderson2012,Bisknell2017,Kelly2017,Fischenich2017} However, a casual survey of the same media stories reveals that most municipalities are driven toward converting by lower total cost of ownership enabled by the improved luminous efficacy of LED luminaires. The rebound effect and impact of shifting the spectrum of light emitted by street lighting systems to short wavelengths can displace the potential benefits of SSL to communities. Claims about the purported benefits of SSL may well be dubious, and a lack of sound research can perpetuate unfounded myths about these benefits.~\citep{Marchant2004,Marchant2005,Marchant2017} To the extent that conversion to SSL results in changes to skyglow over cities, there is a strong need to measure conditions before and after the implementation of LED conversions and identify strategies that successfully ensure they do not aggravate the problem.

Tucson, Arizona (U.S.), elected to reduce lighting levels during the conversion of its municipal street lighting system from a mixture of high-pressure sodium (HPS) and low-pressure sodium (LPS) to 3000 Kelvin correlated color temperature white LED in 2016-2017. The design of Tucson's LED lighting system involved the migration of over $4.45{\times}10^{8}$ fully shielded HPS/LPS lumens to roughly $1.42{\times}10^{8}$ fully shielded 3000 K white LED lumens, resulting in a total lumen reduction of 62.8\%. The maximum illuminance directly beneath each street light at the road surface dropped from 60 lux to 17 lux (-72\%) as HPS lighting was removed and replaced with LED luminaires. The program was undertaken by the City of Tucson in part to help protect the assets of several major professional astronomical observatories located nearby, whose collective impact to the local economy is significant.~\citep{Eller2007}

We obtained an interesting and unique dataset in the Tucson area in 2014 that serves as a point of comparison for skyglow after the 2016-17 LED conversion. The data were part of a student project to inter-compare different methods of characterizing the brightness of the night sky through both direct detection of sky radiance and indirect sensing of sky brightness using visual limiting stellar magnitude estimates. While the project goal was to inter-calibrate different measurement methods, the data set forms a record of sky brightness conditions in and around Tucson in the years just prior to the LED conversion. Further, the data were collected in early summer, when weather conditions are typically most favorable for night sky brightness measurements. New data collected after the LED conversion is complete (or nearly so) may reveal changes in skyglow attributable to the new lighting system, if street lighting comprises a significant component of the city's overall light emission budget.~\citep{Kuechly2012} 

This dataset enables us to address a fundamental question: did the skyglow over Tucson change as the result of reduced lighting levels implemented during the municipal LED street lighting conversion? Any net change would be attributable to a combination of greater molecular scattering, as a consequence of fractionally more short-wavelength light emitted by the new LED system, and lower overall light emission, resulting from the City of Tucson's decision to reduce the number of lumens emitted per City-owned luminaire. Since the molecular content of the lower atmosphere fluctuates only by a few percent, we cannot attribute any net change of skyglow to only molecular scatter.

Aerosols are the only atmospheric constituent that can vary significantly, thus modulating skyglow. There is no doubt that backscatter of light is mostly due to molecular scatter, but this is true only if: (1) the particles are large compared to air molecules; and, concurrently, (2) the number concentration of aerosol particles is several orders of magnitude smaller than that of molecules. The size distribution of particles in urban air is conventionally characterized by three modes. The smallest nucleation mode contains particles sized $<0.1$ $\mu$m and is formed by condensation of hot vapor from combustion sources and from chemical conversion of gases to particles. These particles, or even the smallest fraction of accumulation-mode particles, can affect the backscatter significantly also, because the number concentration of these particles is usually high.~\citep{Deirmendjian1969} We therefore endeavored to obtain night sky radiance measurements under comparable atmospheric conditions in order to reduce the chance that differing turbidity would mask skyglow effects properly attributable to light source changes.

Assuming that (1) the reduction of lumens during the LED conversion outweighed the increased upward light scattering contribution resulting from shifting the spectrum of lighting toward shorter wavelengths, and (2) municipal street lighting accounts for a significant fraction of the overall upward-directed light emissions from Tucson, we expected skyglow to decrease over Tucson as a consequence of the conversion. Further, the expected reduction in skyglow was simply proportional to the reduction in the municipal street lighting emission, adjusted according to the anticipated increase in light scattering. This is because no other changes were made: luminaire mounting height, pole spacing, target albedo and other factors were left unchanged during the conversion. In order to address the research question, we compared the observations with results of radiative transfer model runs describing both the ``before'' and ``after'' conversion scenarios. We also measured change in the upward-directed radiance from the city as seen by Earth-orbiting satellites.

This paper is organized as follows. First, in Section~\ref{sec:simulations}, we describe the radiative transfer model used to predict relative skyglow changes after the completion of the Tucson LED conversion project. Next, in Section~\ref{sec:measurements}, we review the site selection and measurement methods for ground-based skyglow observation campaigns in 2014 and 2017. The results are presented in Section ~\ref{sec:analysis} along with an analysis in the context of our modeling outcomes. Finally, in Section~\ref{sec:summary}, we summarize our work, point out its limitations, and comment on the applicability of the results to other LED conversion efforts.

%
%
\section{SkyGlow Simulator Predictions}\label{sec:simulations}

\subsection{Light clustering approach}
The algorithm we have used to model the sky radiance and luminance distributions for Tucson is based on the theory developed by Kocifaj \citep{Kocifaj2007} and improved in later releases. The software solution ``Skyglow v.5'' is publicly available\footnote{\url{http://skyglow.sav.sk/\#simulator}.} and can be used to simulate sky radiance or luminance patterns over any place in the world. The SkyGlow tool allows for clustering of the light-emitting areas that share similar properties such as spectral power distribution (SPD), average number of lumens installed per unit land area, angular emission pattern, and the spectral reflectance of the ground.

\begin{figure}[tbh]
\centerline{\includegraphics[scale=1.1]{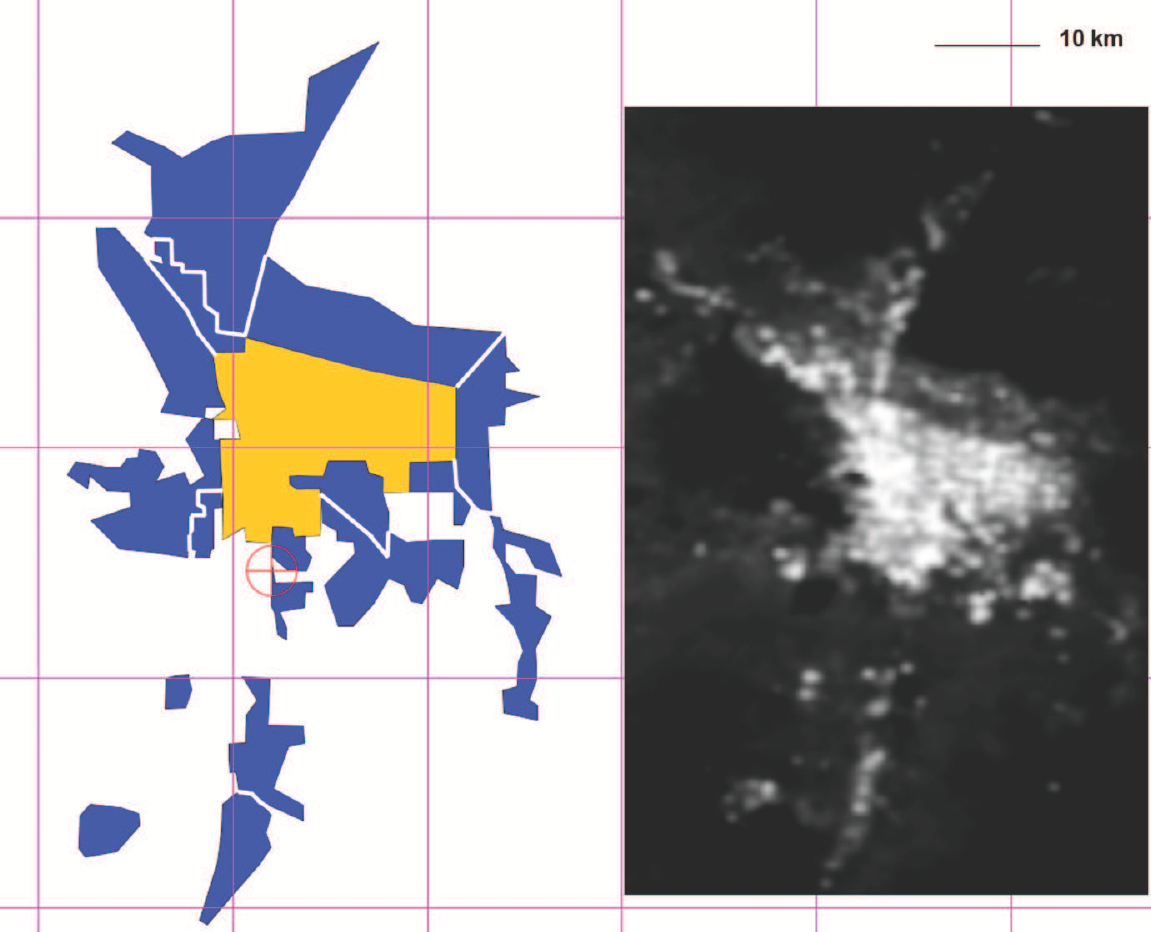}}
\caption{Model of the Tucson city and suburban region illustrating the 18 polygonal modeling units referred to in the text. The inset image at right shows a 2017 NASA/NOAA \emph{Suomi National Polar-orbiting Partnership} Visible Infrared Imaging Radiometer Suite Day-Night Band (VIIRS-DNB) image of the area at night, at approximately the same spatial scale and orientation.}
\label{tucsonmap}
\end{figure}
The City of Tucson was divided into 18 light-emitting areas that share common physical properties (Figure~\ref{tucsonmap}), meaning that, e.g., uplight levels or built-up area densities fall within the same categories. Our analysis was based on Google Earth, Day-Night Band (DNB) maps made using the Visible Infrared Imaging Radiometer Suite (VIIRS) instrument aboard the NASA/NOAA \emph{Suomi National Polar-orbiting Partnership} satellite \citep{Murphy2006}, and NASA ``City Lights'' that is part of Google Earth gallery\footnote{\url{https://www.gearthblog.com/blog/archives/2012/02/the_city_lights_of_earth.html}.}. A complete inventory of luminaires in the central part of Tucson was available for City-owned lights, and a partial inventory for other lights. However, the information on lumens installed in suburban areas was completely missing; it was thus determined numerically and normalized using the VIIRS database. 

The procedure was simple. First we analyzed a few areas in central Tucson, where total emission spectra were computed as the collective contribution of HPS, LPS and LED lamps, taking into account the information on initial lamp lumens and luminaire efficiency. We found that the ratio of VIIRS-DNB uplight to lumens installed varies only slightly for these areas ($\pm$4\%); thus, the same ratio was used to estimate lumens installed in other suburban zones. Such calibration was possible also because the mean ground albedo does not significantly change across the city territory.

Using the luminaire inventory and the above calibration we found that, prior to the LED conversion, the central part of Tucson emitted  $8.52{\times}10^{8}$ photopic lumens, constituting a mixture of 93\% HPS, 5.8\% LPS, and 1.2\% 3000 K white LED. We determined that the legacy street lighting system comprised  $4.81{\times}10^{8}$ lumens, which was 56.4\% of lumens installed from all sources, both public and private. The City of Tucson chose to replace the legacy system with new LED luminaires whose output is 63\% less than the existing lighting.

City planners estimated that the LED lamps would emit  $1.79{\times}10^{8}$ lumens if operated at maximum power, leaving  $3.71{\times}10^{8}$ lumens (43.6\% of all light emissions) unchanged. This results in a total of $5.50{\times}10^{8}$ lumens from all sources, and represents a 35.4\% reduction from the pre-retrofit condition if the new luminaires were operated at full power. However, the City elected to further dim the new LED lights to 90\% of their maximum power upon installation, so the total post-retrofit emission of the city was $5.33{\times}10^{8}$ lumens, for a total reduction of 37.6\% from the pre-retrofit condition.

\subsection{Model lighting scenarios and inputs}

The models in this study comprise two scenarios:

M1: Status quo ante (pre-retrofit condition):  $8.52{\times}10^{8}$ lumens installed prior the LED conversion

\begin{figure}[tbh]
\centerline{\includegraphics[width=1.0\linewidth]{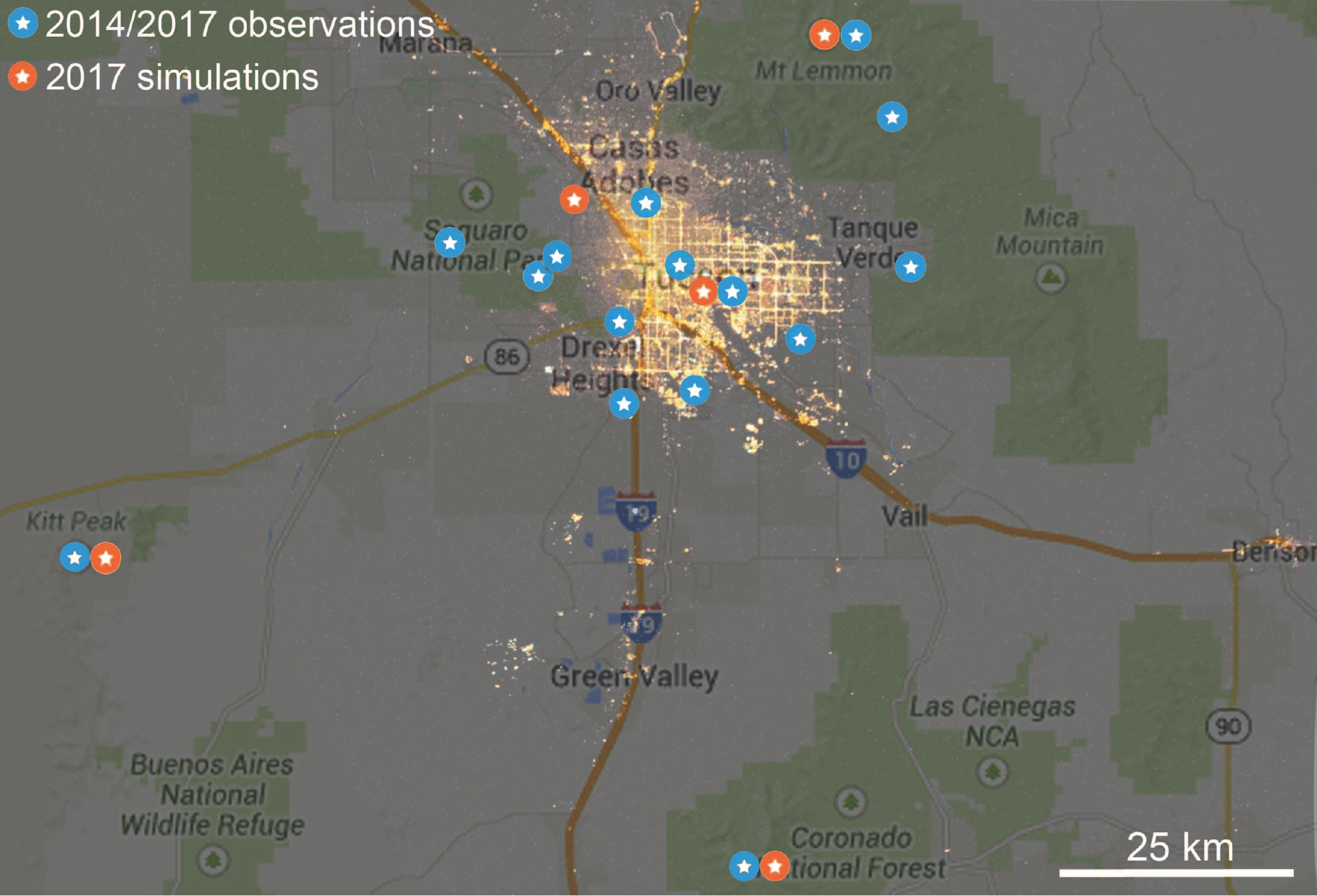}}
\caption{A nighttime optical-light photograph of the Tucson, Arizona, vicinity from low Earth orbit shown superimposed on a background political map of the area; locations of SkyGlow Simulator model runs are indicated with red symbols and 2014/2017 night sky brightness measurements with blue symbols. Light at night image: National Aeronautics and Space Administration Astronaut Photo ISS030-E-61700, obtained on 31 January 2012. Background map: copyright 2017 Google, INEGI, used with permission.}
\label{tucsonsites}
\end{figure}

M2: Post-retrofit, dimming to 90\% output:  $5.33{\times}10^{8}$ lumens after LED conversion (56.4\% lumens undergo conversion, while the lights are dimmed to 90\% of maximum power). The total post-conversion lumen output, $I_{\textrm{post}}$, is obtained from the sum of (90\%-dimmed) street lighting and unchanged, non-street lighting:
\begin{equation} 
I_{\textrm{post}} = (0.9 \times \textrm{$1.79{\times}10^{8}$ lm}) + (0.436 \times \textrm{$8.52{\times}10^{8}$ lm}).
\end{equation}
The best we can do to get closer to at least a partially-controlled experiment is to make both the numerical modeling and field measurements under similar sky conditions (clear sky, low dust, low relative humidity, no moonlight, no twilight, no Milky Way in the zenith). The input parameters to the model were kept constant unless stated otherwise. For instance, the aerosol optical depth (AOD) at the reference wavelength ($\lambda$ = 500 nm) was 0.1, while the \r{A}ngstr\"{o}m exponent was $\upsilon = 1.3$. The latter parameter models AOD as an exponential function of wavelength: AOD ${\sim}$  ${\lambda}^{-\upsilon}$. Models were computed for five locations in and around Tucson, shown in Figure~\ref{tucsonsites}.

\subsection{Results of model runs}
To illustrate results exemplifying the city core, skyglow modeling and baseline/post conversion empirical measurements are indicated for the Reid Park Brown Conservation Learning Center (32$^{\circ}$12$^{\prime}$29.7$^{\prime\prime}$N 110$^{\circ}$55$^{\prime}$17.8$^{\prime\prime}$W; see the red cross in Figure 3). The luminance distribution as well as the percent change when transitioning from M1 to M2 for Reid Park are shown in Figure 4. The zenith luminance and horizontal illuminance computed were approximately 3.5 mcd m$^{-2}$ and 25 mlux, respectively.
\begin{figure}[tbh]
\centerline{\includegraphics[width=0.8\linewidth]{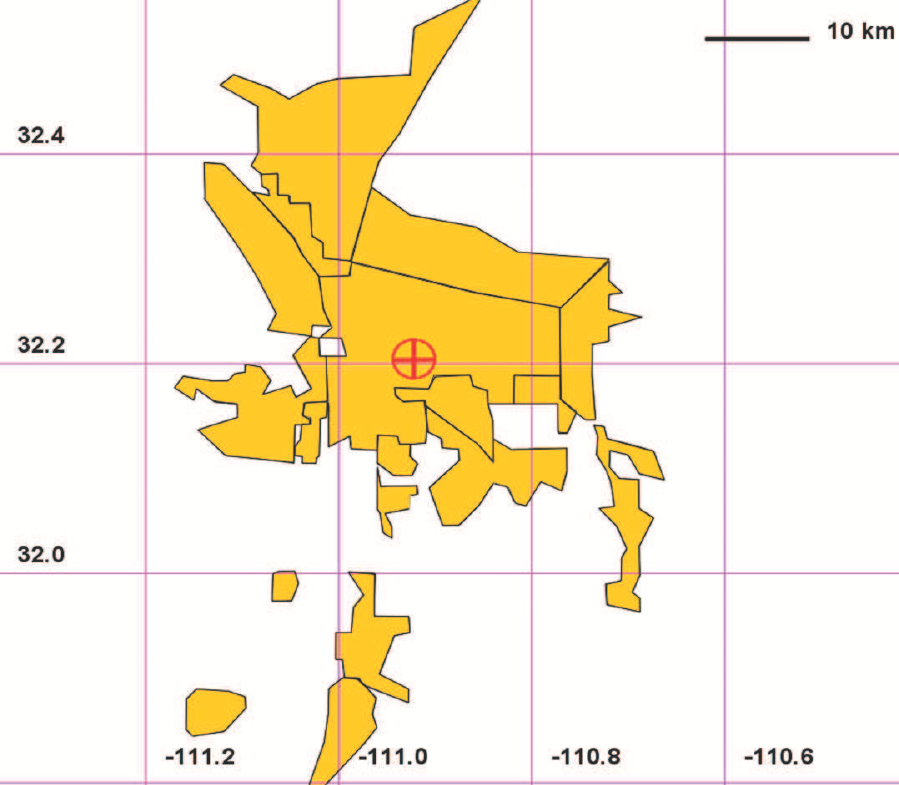}}
\caption{A polygonal map of Tucson with a discrete observation site (Reid Park Brown Conservation Learning Center) marked by the red cross in the red circle. This site represents model predictions typical of the city core.}
\label{polygons}
\end{figure}
\begin{figure}
\centering
\begin{minipage}{.5\textwidth}
  \centering
  \includegraphics[width=.90\linewidth]{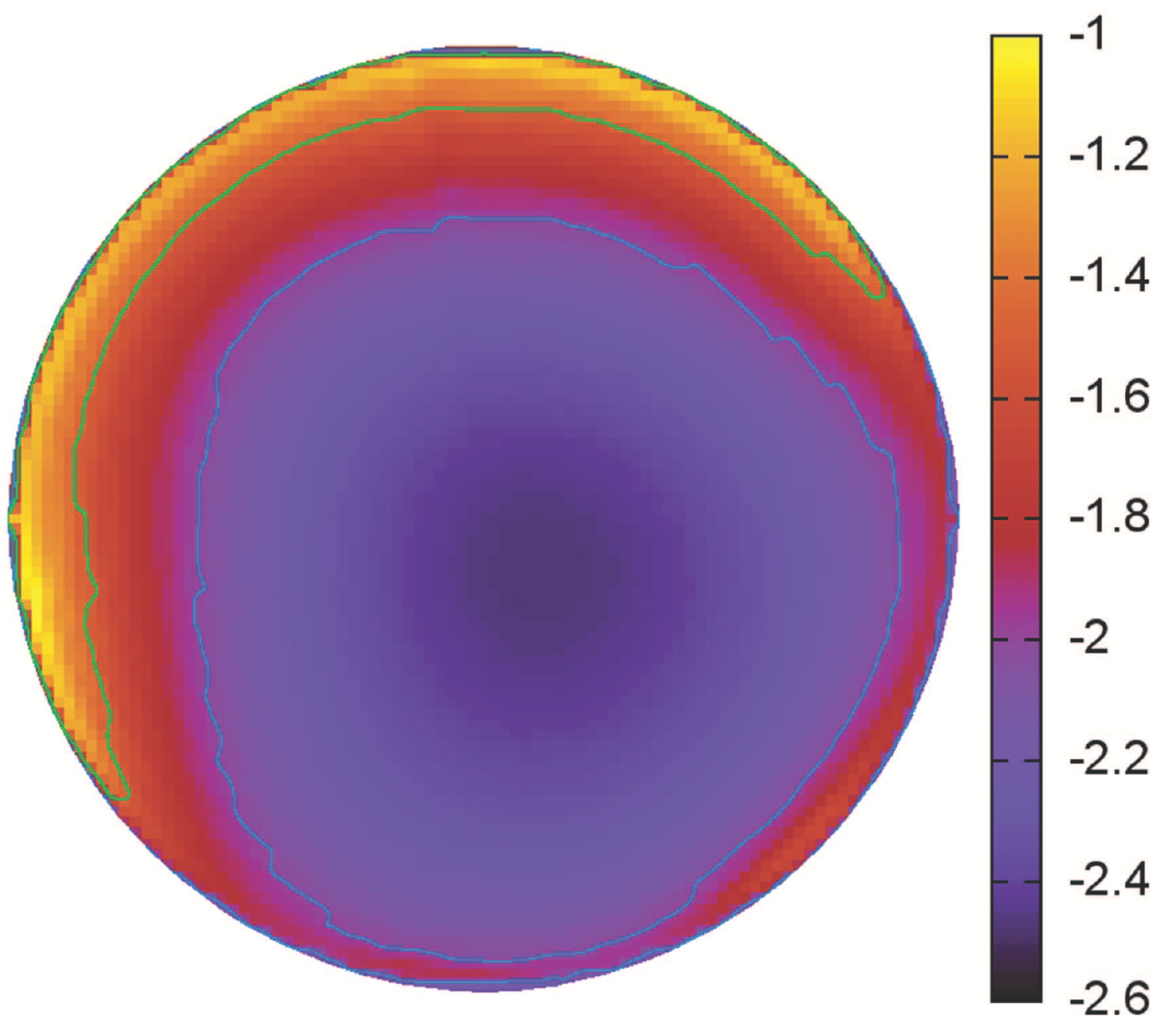}
\end{minipage}%
\begin{minipage}{.5\textwidth}
  \centering
  \includegraphics[width=.90\linewidth]{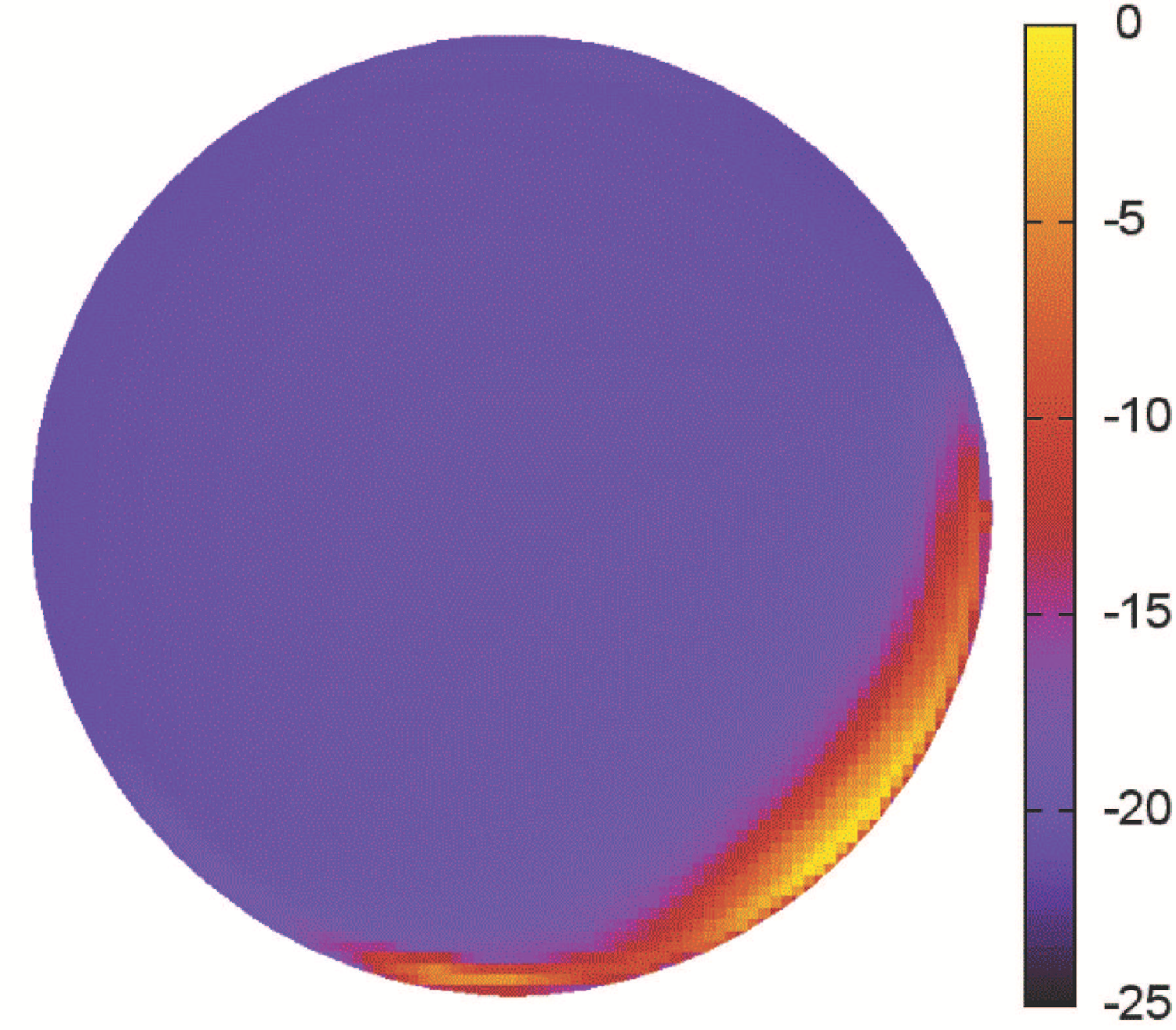}
  \label{fig:test2}
\end{minipage}
\caption{Model outputs for the Reid Park Brown Conservation Learning Center site. Left: All-sky luminance distribution in units of log$_{10}$ cd m$^{-2}$. Right: Percent change in all-sky luminance when transitioning from M1 to M2. North is up and east is at right in both images. Because backscatter dominates over other forms of scattering in and near urban environments, the largest changes in skyglow are seen in directions opposite the azimuthal position of the dominant skyglow source.}
\label{reidparklum}
\end{figure}

Bear in mind that the focus of the model is not on absolute values, but on the relative influence of lumen output change, given the uncertainty of the normalization coefficient for the VIIRS data and the intent of this study to examine skyglow relative to the situation prior to the LED conversion. This is why the rightmost plot in Figure~\ref{reidparklum} and consequent graphical outputs show the percent change only. For example, a value of +20\% for model M2 means a conversion with LED lights dimmed to 90\% increases the sky glow by 20\% or 1.2 times the M1 result, whereas a negative value of -20\% means the sky glow will decrease by 20\% or 0.8 times the M1 value.

\begin{figure}[!p]
\centerline{\includegraphics[width=1.0\linewidth]{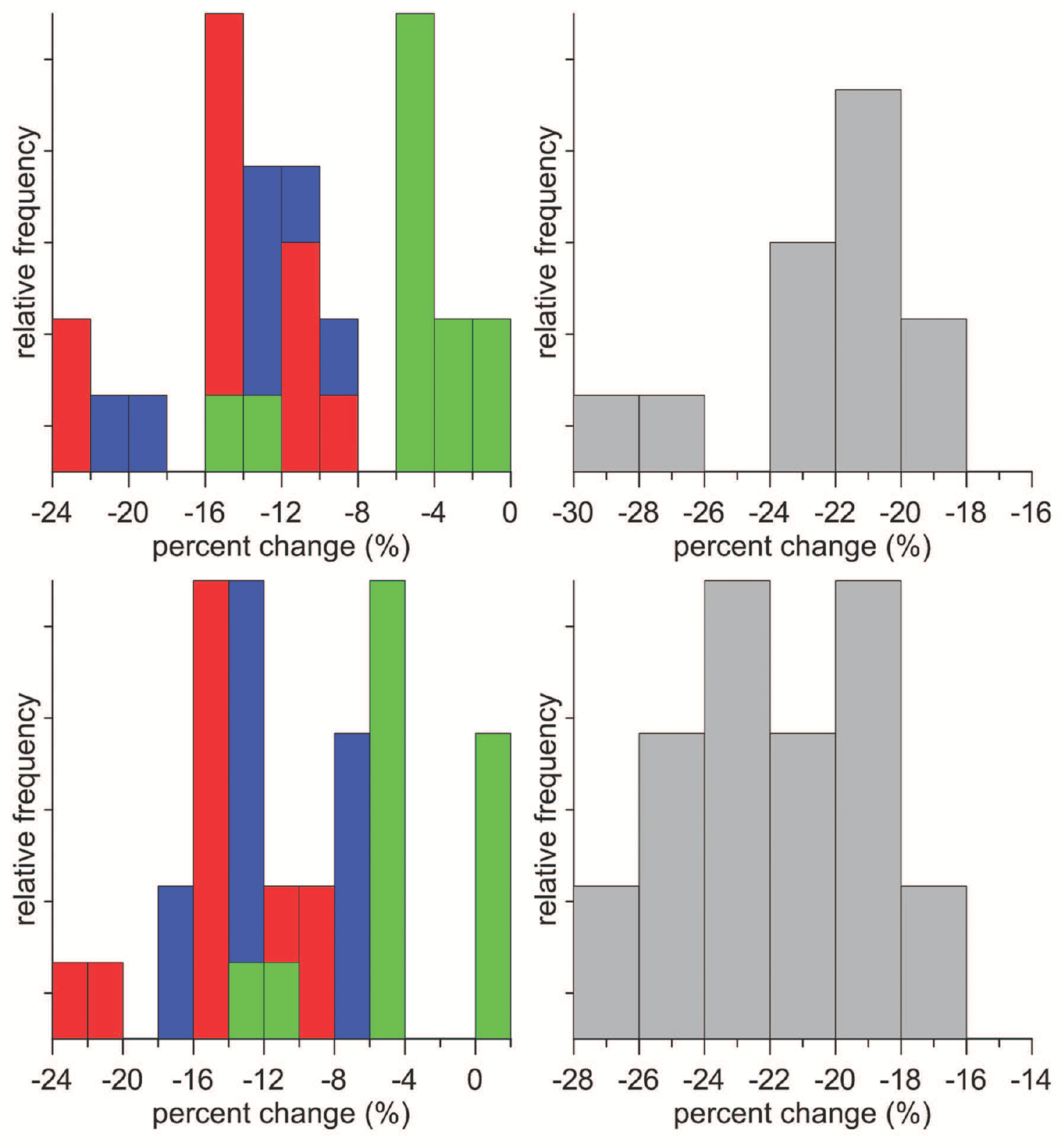}}
\caption{The percent change of skyglow in the transition from M1 to M2. M1 represents the theoretical sky state in Reid Park prior LED conversion, while M2 is for sky after conversion with LED lights dimmed to 90\% of maximum power. The plots at top left are for diffuse horizontal irradiance in the blue, green and red parts of the spectrum. Diffuse scotopic illuminance predictions are shown at top right. The bottom panels are for zenith radiance in the blue, green and red (bottom left), and zenith luminance (bottom right).}
\label{reidparkbars}
\end{figure}

The percent change was computed for every radiance/luminance or irradiance/illuminance. A change of -20\% was predicted for almost all computed luminance data for Reid Park, while no change is only observed in the part of the sky opposite to the azimuthal position of the dominant light source; compare the left and right plots in Figure~\ref{reidparkbars}, which shows the percent changes predicted by the models for the transition from M1 to M2. Here we have assumed that fully shielded LED lights are mixed with lights that are not properly shielded, implying that the direct uplight ranges from 0\% to 5\%.

Computations were made for different combinations of $F$ and AOD values in order to identify the statistical range of the optical effects we studied. There is therefore no reason a priori to expect that the red bars in Figure~\ref{reidparkbars} will be completely isolated from green bars and blue bars. Instead, a partial overlap of blue, green and red bars is seen in the figure. Additionally, the arrangement of blue, green and red bars in the figure is due to a non-trivial combination of different optical effects; e.g., spectral power distribution and atmospheric optics, including an intensive scattering in the blue, but also an elevated value of AOD in blue. While a large AOD implies more scattering events and increased scattering efficiency, it also means more rapid intensity decay because of increased extinction. AOD is low for red wavelengths, so the scattering efficiency and extinction are both low. The order in which the colored bars appear in Figure~\ref{reidparkbars} does not depend on the spectral arrangement of the optical effects studied, since the computations were made under differing combinations of $F$ and AOD.

The spectral irradiances before and after the LED conversion change only slightly with AOD. This is because backscatter typically dominates other effects when forming skyglow in the territory of a city or in its vicinity. We know from light scattering theory that backscatter from aerosol particles is typically low compared to side-scatter or forward scatter, so solid and liquid particles are the strongest modulators of ground-reaching radiation in distant places. However, the effect that aerosols have on skyglow is lowered if the horizontal distance to a light source is small, i.e.,~when large scattering angles become decisive in forming the diffuse light of a night sky.  

In the example shown in the left panel of Figure~\ref{reidparklum}, the dominant light sources are situated along directions ranging from northeast to southwest along the horizon. The percent change in luminance is shown in the right panel, where the maximum values are seen toward the southeast. The maximum value in terms of percent change is near zero, whereas the minimum value of -20\% is seen over almost all of the rest of the sky. Note that the percent change is not an amplitude, and that the sign of the change is therefore important. Since we expect that the luminance decrease resulting from changes to the whole-city light emission in the LED conversion affects the entire sky, the percent change in luminance in the direction of backscatter is close to zero (-20\% + 20\% = 0\%), explaining the perhaps counterintuitive appearance of the figure.

When cities are large in geographic extent ($>$20-30 km) and the distance between the light source and observer is comparable to the city size, the dominant source of ground-reaching irradiance is the fraction of light radiated directly into the upward hemisphere, $F$. For Reid Park, the percent change in clear-sky spectral irradiance was predicted to increase by a factor of two when transitioning from $F=0$\% to $F=5$\%. However, an overcast sky has the opposite behavior, yielding a decrease of spectral irradiance by a factor of two (not shown here). This is because for $F=0$\%, photons are effectively ``trapped'' in city territory where backscattering dominates the formation of skyglow, while photons emitted at small upward angles reach the cloud level at relatively large horizontal distances and therefore do not increase irradiance in the city. 

Although all LED lamps were assumed to emit only downwards, we have also introduced 5\% uplight into the results presented in Figure~\ref{reidparklum} in order to simulate combined effect of fully shielded lights and incumbent luminaire distributions. The 5\% uplight figure was chosen following the estimate of Kinzey et al.~\citep{Kinzey2017} for ``relatively poor drop-lens cobra head street lights.'' While $F=0$\% and low AOD is the condition most representative of the field skyglow measurements reported here, we varied both parameters.

\begin{figure}[tbh]
\centerline{\includegraphics[scale=0.72]{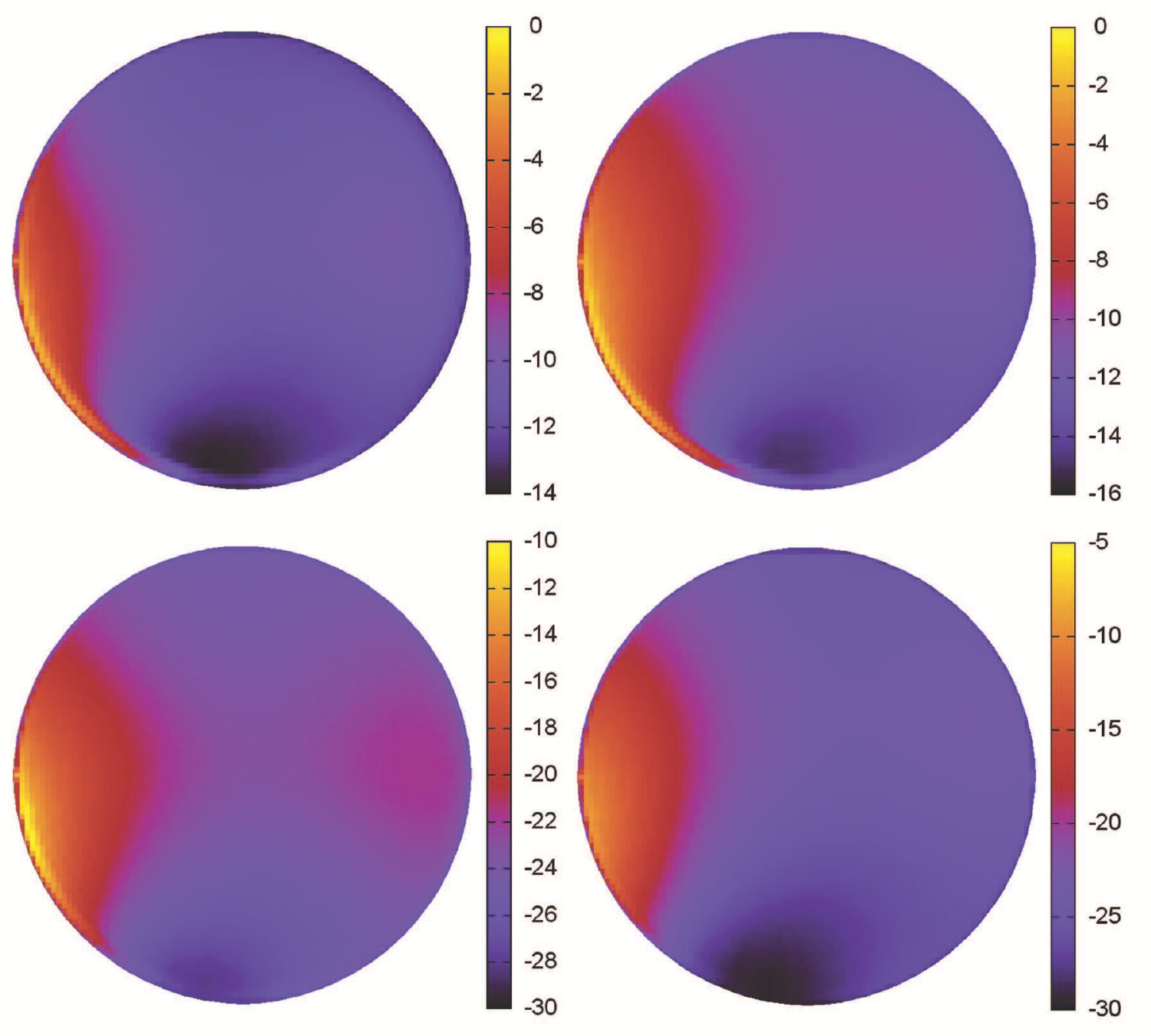}}
\caption{Predicted percent change in all-sky luminance at the Mount Lemmon SkyCenter when transitioning from M1 to M2. The panels at the top are for fully shielded lights (F=0\%), while the bottom panels are for F=5\%. From left to right: AOD$_{500}=0.1$, and 0.5, where AOD$_{500}$ is the aerosol optical depth at the nominal wavelength of 500 nm. North is up and east is at right in all images.}
\label{mtlemmonlum}
\end{figure}
The maximum percent change for skyglow for the Mount Lemmon SkyCenter (32$^{\circ}$26$^{\prime}$32.1$^{\prime\prime}$N 110$^{\circ}$47$^{\prime}$20.0$^{\prime\prime}$W; uppermost symbols in Figure~\ref{tucsonsites}) is shown in Figure 6. The SkyCenter represents a sensitive astronomical observatory facility beyond the territory of the City of Tucson and its suburbs. Model results consider both low and high atmospheric turbidity conditions assuming direct uplight is either zero (the plots at the top of Figure~\ref{mtlemmonlum}) or $F=5$\% (the bottom panels). The computational results for Mount Lemmon indicate the effect of AOD is apparent especially for $F=0$\%. A negative percent change in sky luminance is found at small angular distances from a light source if AOD is low. In a turbid atmosphere with AOD as large as 0.5, the percent change is small near the azimuthal position of the dominant light source and decreases only if angular distance increases (compare left and right plots at the top of Figure~\ref{mtlemmonlum}). 

Unlike Reid Park, Mount Lemmon is outside the city territory, implying that side scatter is equally or even more important than backscatter in forming the skyglow over such sites. Therefore, the turbidity increase is reflected in the percent change in luminance for only some parts of sky. In the case of fully shielded luminaires, the percent change in luminance is roughly two times smaller than that for $F=5$\% (compare the plots at the top and bottom in Fig. 6). This result coincides well with what we have found for other places toward the urban core of Tucson, including Reid Park.

%
%
\section{Skyglow Measurements}\label{sec:measurements}

The field measurements reported in this paper resulted from an effort, carried out in June 2014, to inter-compare various approaches to characterizing the brightness of the night sky. The goal of this project was to assess the reliability of different methods in relation to one another, and to look for any systematic biases in the measurement approaches or dependencies among them. In 2016, we realized that this data set comprised a useful assessment of skyglow conditions in and around Tucson before the municipal LED lighting conversion began. We repeated the 2014 observations in May and June 2017, anticipating completion of the conversion project by 1 May 2017. In reality, the conversion was completed around 1 August; at the time of our second-epoch observations, the conversion was approximately 95\% complete.

\subsection{Site selection}
We used a variety of data sources to characterize the visual brightness of the night sky from 15 locations in and around Tucson; the locations and visit dates of each place included in the sample are listed in Table~\ref{skyglowsites}. The measurement sites are also shown on the map in Figure~\ref{tucsonsites}. The selection of measurement sites intended to probe both urban conditions within the continuously-built environment of the Tucson area, as well as a number of more distant locations whose skies are impacted by the Tucson light dome. In each case, the geospatial distribution of measurement locations was considered, as well as the relative ease of nighttime access. In some instances, and particularly for the suburban/rural sites, we chose astronomical observatories with histories of their own measurements of night sky brightness.
\begin{sidewaystable}
\medskip
\resizebox{\linewidth}{!}{%
\begin{tabular}{ lccccc }
 \hline
 \textbf{Site name} & \textbf{Lat (deg)} & \textbf{Long (deg)} & \textbf{Elevation (m)} & \textbf{Visit 1} & \textbf{Visit 2}\\
 \hline
 \it Cooper Center for Environmental Learning \rm & +32.24119 & $-$111.08097 & 859 & 6/18/2014 & 5/23/2017\\
 \it Fred Lawrence Whipple Observatory \rm & +31.68060 & $-$110.87886 & 2345 & 6/24/2014 & 6/3/2017\\
 Gates Pass & +32.22341 & $-$111.10156 & 964 & 6/18/2014 & 5/23/2017\\
 John F. Kennedy Park & +32.18135 & $-$111.01299 & 753 & 6/22/2014 & 5/22/2017\\
 \it Kitt Peak National Observatory \rm & +31.96072 & $-$111.59949 & 2080 & 6/25/2014 & (Note 1)\\
 \it Mission San Xavier del Bac \rm & +32.10644 & $-$111.00787 & 767 & 6/19/2014 & 5/22/2017\\
 Mt. Lemmon SkyCenter & +32.44225 & $-$110.78889 & 2789 & 6/23/2014 & 5/28/2017\\ 
 \it National Optical Astronomy Observatory Headquarters \rm & +32.23346 & $-$110.94720 & 747 & 6/17/2014 & (Note 2)\\
 \it Pima Community College East Campus \rm & +32.16586 & $-$110.81629 & 850 & 6/20/2014 & 6/16/2017\\
 \it Reid Park Brown Conservation Learning Center \rm & +32.20826  & $-$110.92161 & 757 & 6/21/2014 & 5/21/2017\\
 \it Saguaro National Park Rincon Mountain District \rm & +32.17992 & $-$110.73643 & 941 & 6/20/2014 & 6/16/2017\\
 \it Saguaro National Park Tucson Mountain District \rm & +32.25366 & $-$111.19744 & 779 & 6/22/2014 & 5/22/2017\\
 \it Tucson Auto Mall \rm & +32.28974 & $-$110.98444 & 701 & 6/21/2014 & 5/21/2017\\
 \it Tucson International Airport \rm & +32.11841 & $-$110.93128 & 793 & 6/19/2014 & 6/16/2017\\
 \it University of Arizona Campus \rm & +32.23189 & $-$110.94665 & 747 & 6/17/2014 & 5/21/2017\\
 \it Windy Point \rm & +32.36827 & $-$110.71675 & 2002 & 6/23/2014 & 5/28/2017\\
 \hline
\end{tabular}}
\caption{Summary of locations in and near Tucson, Arizona, where sky brightness data were obtained in 2014 and 2017. The dates of pre- and post-LED conversion visits are given in columns 5 and 6, respectively. Sites that did not yield useful data, or did not receive a follow-up visit in 2017 under good observing conditions, are indicated by name \emph{in italics}. Notes in column 6: (1) we were unable to make a repeat visit under good weather conditions before the onset of the 2017 summer monsoon season; (2) we decided not to make a repeat visit to this site because of its proximity ($\sim$200 m) to the University of Arizona Campus site.}
\label{skyglowsites}
\end{sidewaystable}

\subsection{Data sources and acquisition}
We used various methods of characterizing the brightness of the night sky as summarized in Table 2. The direct luminance data sources included four narrow-angle Sky Quality Meter (``SQM-L'') devices \citep{Cinzano2007}, the Dark Sky Meter iPhone app \citep{DSM}, and luminance-calibrated all-sky digital imagery \citep{Kollath2017}. Estimates of the naked-eye limiting magnitude (NELM) were made at each site using both the Loss Of The Night app for Android \citep{Kyba2015} as well as the Globe At Night reference charts \citep{GANCharts}. The SQM-L units we used were serial numbers 3829, 5428, 5442, and 8161. While we collected limiting visual stellar magnitude estimates, we found that they varied too greatly between observers to provide a reliable indirect measurement of the sky luminance.
\begin{table}[tbh]
\medskip
\resizebox{\linewidth}{!}{%
\begin{tabular}{ lll }
 \hline
 \textbf{Source} & \textbf{Measurement} & \textbf{Reference}\\
 \hline
 Sky Quality Meter with lens (SQM-L) & Single-channel broadband radiance & \citep{Cinzano2007} \\
 Globe At Night & Naked-eye limiting magnitude estimate & \citep{Kyba2013,Birriel2014} \\
 Loss Of The Night app & Naked-eye limiting magnitude estimate & \citep{Kyba2015} \\
 Dark Sky Meter app & Spatially-averaged, multichannel illuminance & \citep{DSM} \\ 
 Digital single-lens reflex (DSLR) camera & Spatially-resolved multichannel illuminance & \citep{Kollath2017} \\
 \hline
\end{tabular}}
\caption{Summary of data sources used to characterize visual night sky brightness in this study.}
\label{measurements}
\end{table}

We endeavored as much as possible during the measurements to avoid contamination from local sources of glare, and to account for the presence of the Milky Way in the measurements. Some observations in 2014 were made in conditions that were neither astronomically dark nor under fully clear skies. We noted these anomalies in our analysis and rejected any measurements known to have been thusly compromised.

The native measurement unit of the SQM-L device is the magnitude per square arcsecond (mag arcsec$^{-2}$), a non-SI unit mostly used by astronomers. An approximate conversion between mag arcsec$^{-2}$ and the SI unit of surface brightness, the candela per square meter, is given in \citep{Kubala2009} as:
\begin{equation}
S \textrm{(cd m$^{-2}$)} = (1.08\times10^{5}) \times 10^{-0.4 \times S\textrm{(mag arcsec$^{-2}$)}}.
\end{equation}
Because the photometric passband of the SQM-L differs from the photopic vision response curve, the SQM-L output is a device-specific, spectrally weighted broadband radiance and not properly a luminance. However, because its passband is similar to the photopic curve, SQM-L measurements can be considered approximate luminances. 

SQM-L measurements were made using handheld devices aimed at the zenith. We discarded the initial reading in each set, which tends to be systematically brighter or darker than others in a series. This is a known issue with the device thought to result from slight internal heating of the sensor when power is initially applied.~\citep{Tekatch2017} We then took and recorded at least five measurements in the direction of the zenith in quick succession. In the 2014 campaign, five readings were obtained at each location; during the 2017 campaign, we took 30 measurements at each location for improved statistics and to look for any systematic trends.

For the all-sky imagery, we used an off-the-shelf Canon T5i DSLR body and a Sigma 4 mm circular fisheye lens, giving an apparent field of view of 180 degrees. For consistent orientation of the resulting frames, the camera was pointed at the zenith with the bottom of the camera body oriented toward the northern horizon. The camera settings for all light frames were 30-second exposures at f/2.8 and ISO 1600. The image sequence was: dark, light, light, light, light, light, dark. Dark frames were obtained with the same settings but with the lens cap on, which were later subtracted from light images to remove the contribution from thermal noise in the camera electronics. We allowed the camera body to equilibrate to the ambient air temperature before recording images. No flat fields or other calibration data were obtained.

\subsection{Data reduction}
Different SQM-L units were used during the 2014 and 2017 campaigns. To check the reliability of comparisons between the units, we measured the responses of three of the units (serial numbers 5428, 5442, and 8161) to light under a large range of luminance levels by making simultaneous measurements of the zenith sky during evening twilight on 31 May 2017. The observations were made in non-photometric conditions, which does not affect the validity of the inter-comparison because each of the measurements was made simultaneously with all three devices.

\begin{figure}[tbh]
\centerline{\includegraphics[scale=0.92]{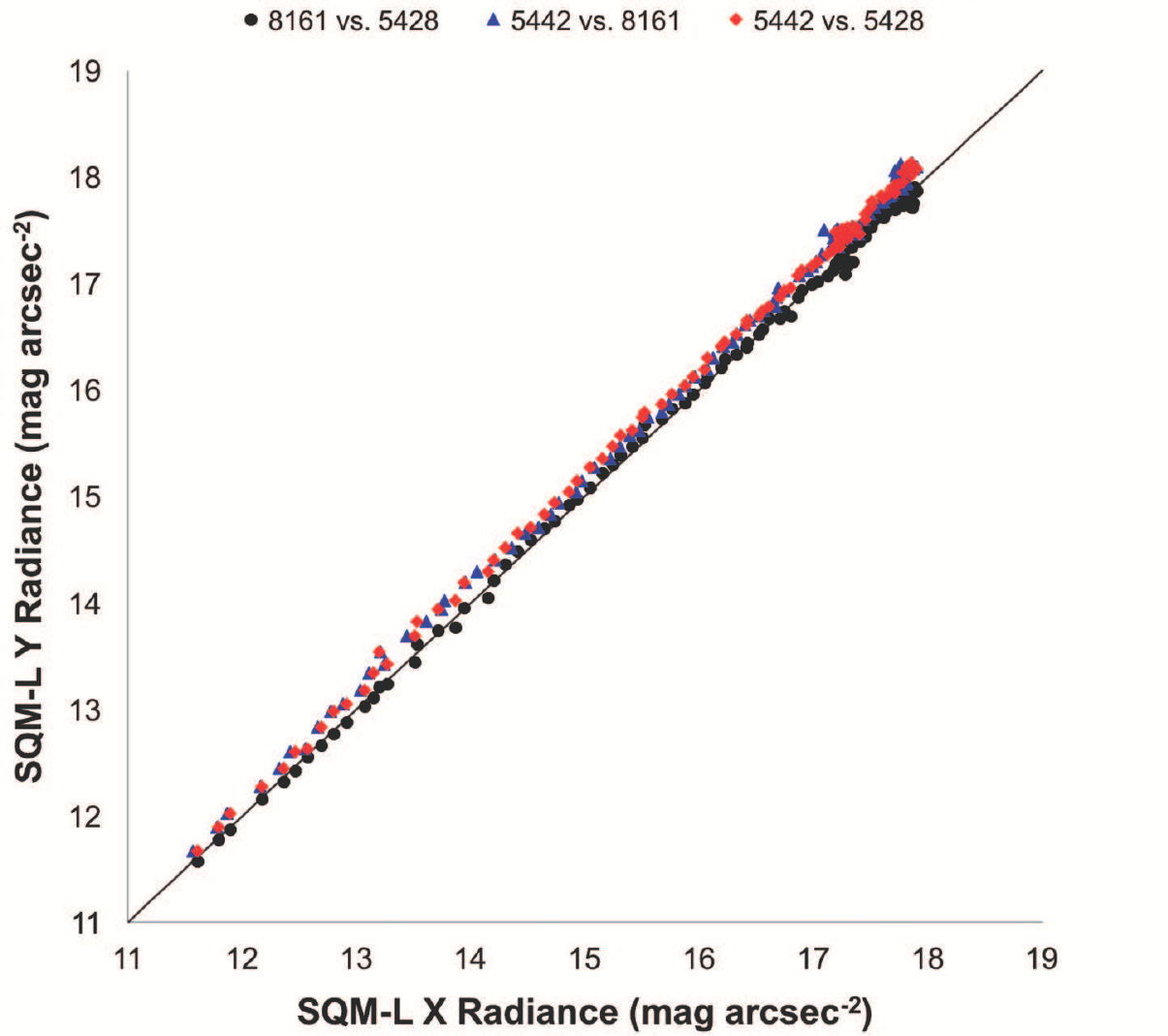}}
\caption{Results of an inter-comparison between SQM-L serial numbers 5428, 5442 and 8161 using as a light source the twilight sky after sunset on 31 May 2017. Each color represents a pair of devices in `Y vs.~X' format: device 8161 vs.~device 5428 (black circles), device 5442 vs.~device 8161 (blue triangles), and device 5442 vs.~device 5428 (red circles). Uncertainties on the individual measurements, $\sim\pm4$\% of the values, are of the order of the symbol size, and thus are not shown. A 1:1 correspondence between the two axes is indicated with a solid black line.}
\label{sqmintercomp}
\end{figure}
Results from the inter-comparison are shown in Figure~\ref{sqmintercomp}.
We derived the following transformation equations for the three SQM-L devices from linear least-squares fits to these data, for broadband radiance $R$ given in mag arcsec$^{-2}$:
\begin{align} 
R(5442) &= (0.0495\pm0.0033) + [(1.0089\pm0.0537) \times R(8161)]\\
R(5442) &= (0.0703\pm0.0027) + [(1.0074\pm0.0429) \times R(5428)]\\
R(5428) &= (0.0146\pm0.0033) + [(1.0011\pm0.0540) \times R(8161)].
\end{align}
We find no large systematic discrepancies between the units when operated under identical conditions of ambient temperature and target luminance. The individual devices showed internal scatter and repeatability comparable to results reported by other authors.~\citep{denOuter2015} The linearity seems to be preserved to lower illuminances, although the fits are less reliable as the data become increasingly noisy. Equations 2-4 are simple linear fits, stated with fitting uncertainties on the parameters. 

The equations enabled us to put measurements from all of the SQM devices we used onto a common photometric system so that meaningful comparisons can be made. We transformed all measurements made with SQM serial numbers 5428 and 8161 to the system defined by serial number 5442. We chose 5442 as the reference since, among the four devices we used, it showed the smallest internal scatter ($\pm$0.03 mag arcsec$^{-2}$) in repeatability tests. Note that SQM-L serial number 3829 was unavailable for use during the 2017 measurement campaign, so we did not include it in the inter-comparison.

All-sky luminance images were calibrated using the method and `dslrlum' software described by Koll\'{a}th and D\"{o}m\'{e}ny \citep{Kollath2017}. The routines read the camera RAW-formatted images, apply spatial distortion/vignetting and luminance corrections, and output several data products. These include a luminance-calibrated version of the input image in cd m$^{-2}$; a Mercator-projected version of this image; and predicted SQM-L values in both mag arcsec$^{-2}$ and cd m$^{-2}$. The predictions are based on photometry of the calibrated images within software apertures of equivalent fields of view; we refer to these values hereafter as ``synthetic'' luminances. The photometry was tied to lab calibration of the camera and lens combination and not to standard stars or other field calibrators. Comparison with actual SQM-L measurements at each site was made as a check, and in most cases the results agreed to within their respective internal uncertainties.

%
%
\section{Analysis}\label{sec:analysis}

A summary of the results of the 2014 and 2017 measurement campaigns is presented in Table~\ref{skyglowresults}. For this analysis we used only the actual and `synthetic' (from all-sky imagery) zenith SQM-L values, being the most reliable figures in the entire data set. We discarded most of the NELM estimates, finding that the values we obtained varied according to observer experience and individual visual acuity too inconsistently in order to draw any reliable conclusions. Instead, we chose to focus on direct luminance measurements. A selection of the calibrated all-sky images, comparing the observations epochs and circumstances in 2014 and 2017, is shown in Figure 8.
\begin{sidewaystable}
\medskip
\resizebox{\linewidth}{!}{%
\tabcolsep=2pt
\begin{tabular}{ |b{9cm}|b{2cm}|b{2cm}|b{2cm}|b{2cm}|b{2cm}| }
 \hline
 \textbf{Location Name} & \textbf{Distance (km)} & \textbf{2014 SQM-L (mag arcsec$^{-2}$)} &  
 \textbf{2017 SQM-L (mag arcsec$^{-2}$)} & \textbf{2014 sSQM-L (mag arcsec$^{-2}$)} & \textbf{2017 sSQM-L (mag arcsec$^{-2}$)} \\
 \hline
 Cooper Center for Environmental Learning & 10.5 & 20.43$\pm$0.02 & 20.46$\pm$0.01 & 19.79$\pm$0.01 & 20.20$\pm$0.01 \\
 Fred Lawrence Whipple Observatory & 61.2 & 21.52$\pm$0.03 & 21.45$\pm$0.02 & 21.54$\pm$0.01 & 21.28$\pm$0.01 \\
 Gates Pass & 12.2 & 20.64$\pm$0.10 & 20.51$\pm$0.02 & 20.20$\pm$0.01 & 20.43$\pm$0.01 \\
 John F. Kennedy Park & 6.1 & 19.61$\pm$0.02 & 19.40$\pm$0.02 & 19.08$\pm$0.01 & 19.07$\pm$0.01 \\
 Mission San Xavier del Bac & 13.1 & 20.02$\pm$0.02 & 20.01$\pm$0.03 & 19.51$\pm$0.04 & 19.82$\pm$0.01 \\
 Mt.~Lemmon SkyCenter & 29.8 & 21.14$\pm$0.08 & 21.38$\pm$0.01 & 21.10$\pm$0.01 & 21.31$\pm$0.01 \\
 Pima Community College East Campus & 16.1 & 19.77$\pm$0.02 & 19.84$\pm$0.04 & 18.99$\pm$0.01 & 19.56$\pm$0.01 \\
 Reid Park & 4.8 & 18.69$\pm$0.06 & 19.01$\pm$0.02 & 18.37$\pm$0.01 & 18.77$\pm$0.01 \\
 Saguaro National Park Rincon Mountain District & 26.9 & 20.93$\pm$0.03 & 20.82$\pm$0.03 & 18.99$\pm$0.01 & 20.74$\pm$0.01 \\
 Saguaro National Park Tucson Mountain District & 21.6 & 20.79$\pm$0.07 & 20.83$\pm$0.02 & 20.50$\pm$0.01 & 20.77$\pm$0.01 \\
 Windy Point & 29.0 & 21.60$\pm$0.07 & 21.36$\pm$0.01 & 18.38$\pm$0.01 & 21.19$\pm$0.01 \\
 Tucson International Airport & 12.2 & 19.14$\pm$0.05 & 19.19$\pm$0.02 & 18.58$\pm$0.02 & 18.74$\pm$0.01 \\
 University of Arizona Campus & 2.5 & 17.50$\pm$0.21 & 18.39$\pm$0.03 & 17.81$\pm$0.01 & 18.38$\pm$0.01 \\
 \hline
\end{tabular}}
\caption{Zenith luminance measurements from the 15 survey locations in 2014 and 2017. ``Distance'' means the radial distance from downtown Tucson (32$^{\circ}$13$^{\prime}$19.3$^{\prime\prime}$N 110$^{\circ}$58$^{\prime}$07.0$^{\prime\prime}$W). For SQM-L data, ``sSQM-L'' refers to `synthetic' SQM-L values derived from aperture photometry of luminance-calibrated all-sky imagery; see the main text for details. Data are shown only where all of our quality criteria were met. Two sites are not shown: No data were obtained at Kitt Peak National Observatory in 2017, and the 2014 Tucson Auto Mall data were not usable.}
\label{skyglowresults}
\end{sidewaystable}

\begin{figure}[H]
\centering
\includegraphics[width=.95\linewidth]{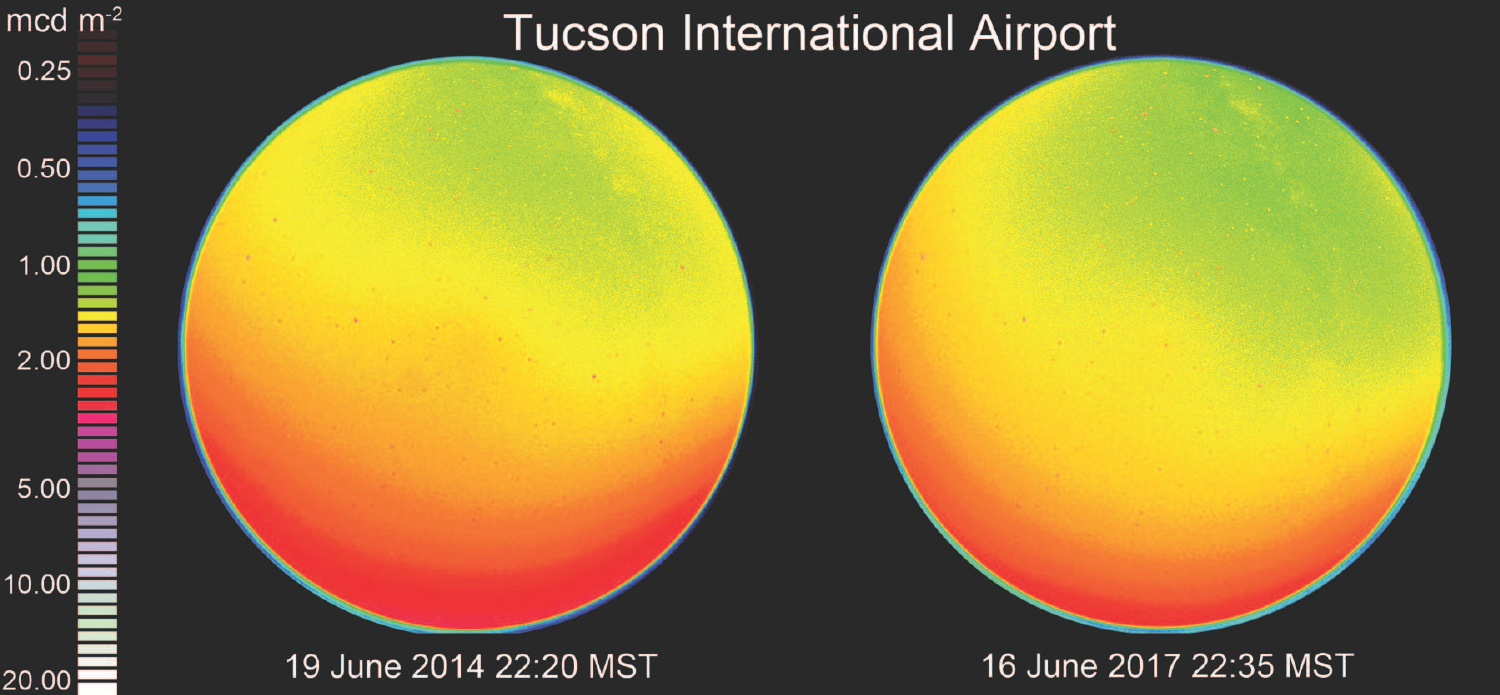}
\includegraphics[width=.95\linewidth]{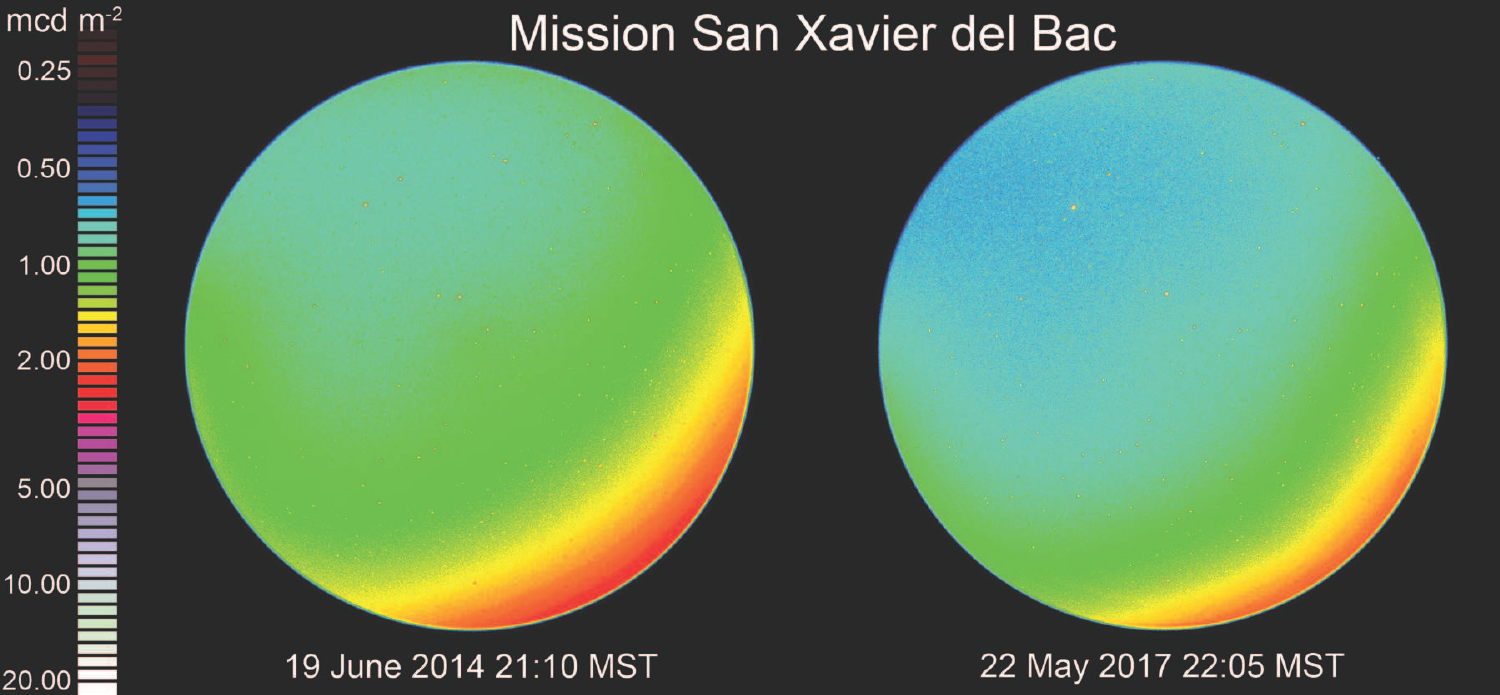}
\includegraphics[width=.95\linewidth]{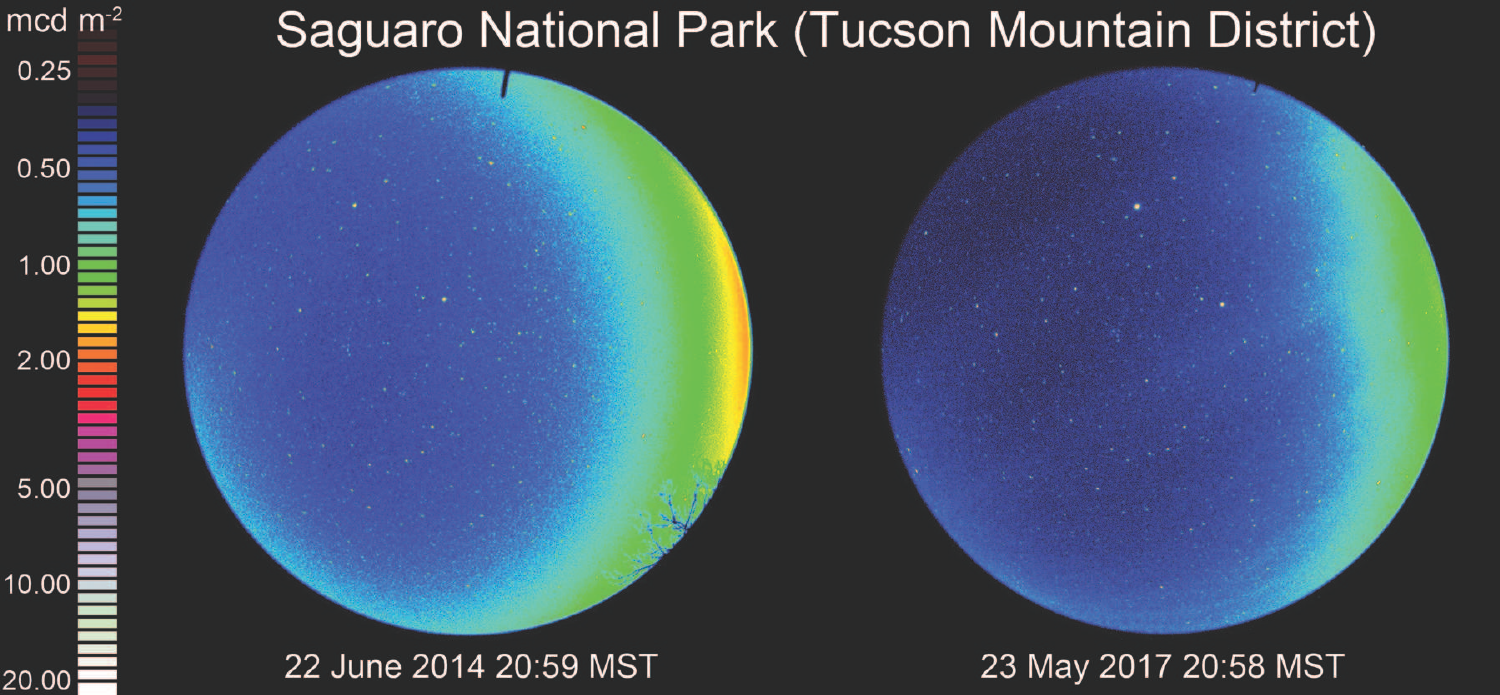}
\caption{Luminance-calibrated all-sky imagery for a selection of sites in our measurement sample. The pre-LED-conversion condition is shown at left in each pair of images, while the post-conversion condition is shown at right. Only sites are shown for which imagery obtained at both epochs was available and met all of our quality criteria, and the images pairs are arranged in order of increasing site distance from downtown Tucson. All images are oriented alike with north down and east at right.}
\label{allskylums}
\end{figure}
\begin{figure}[H]
\centering
\ContinuedFloat 
\includegraphics[width=.95\linewidth]{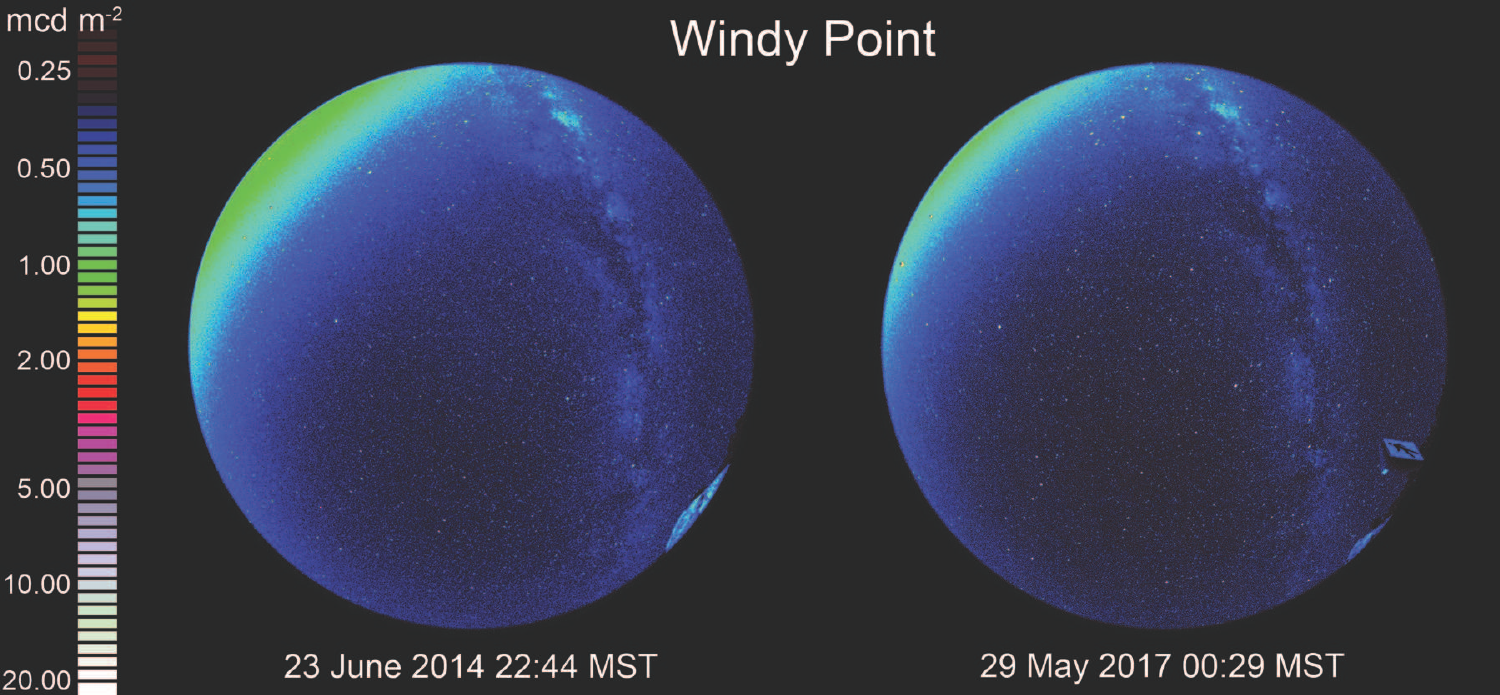}
\includegraphics[width=.95\linewidth]{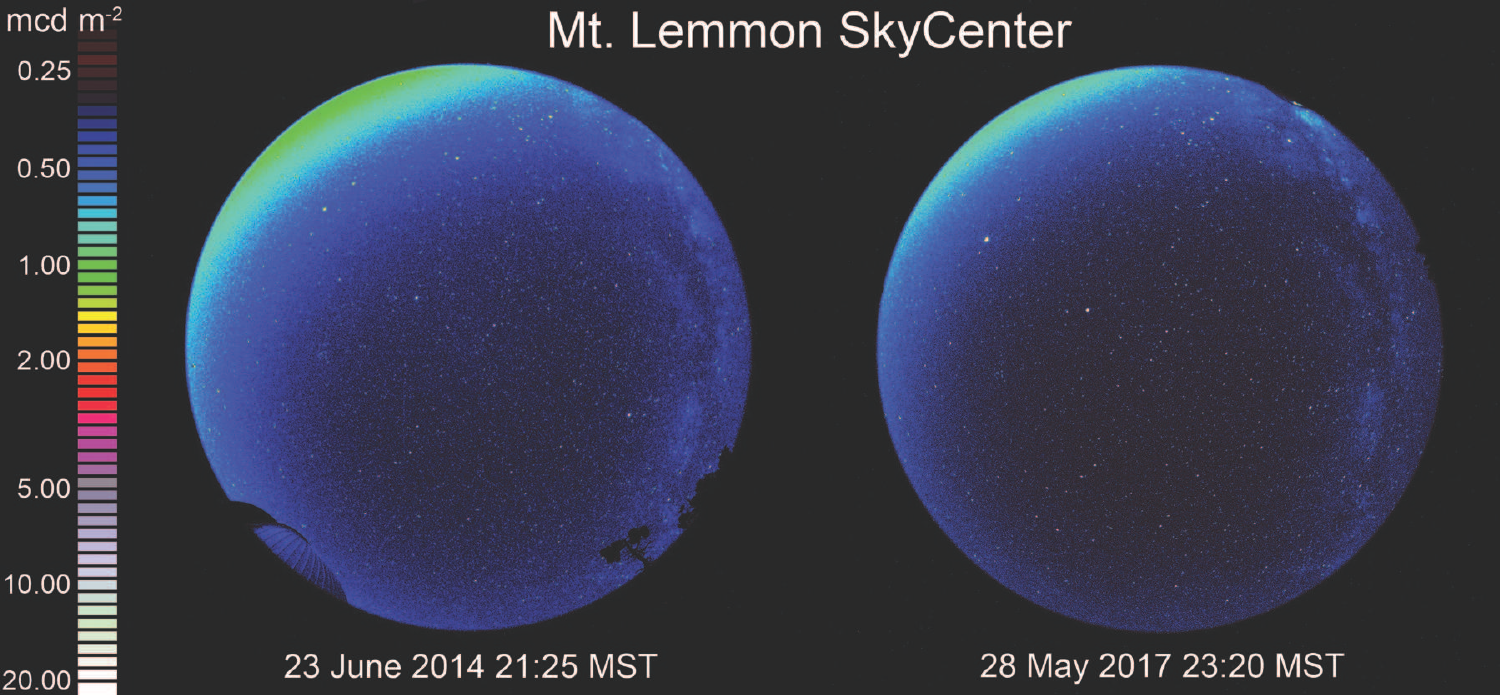}
\includegraphics[width=.95\linewidth]{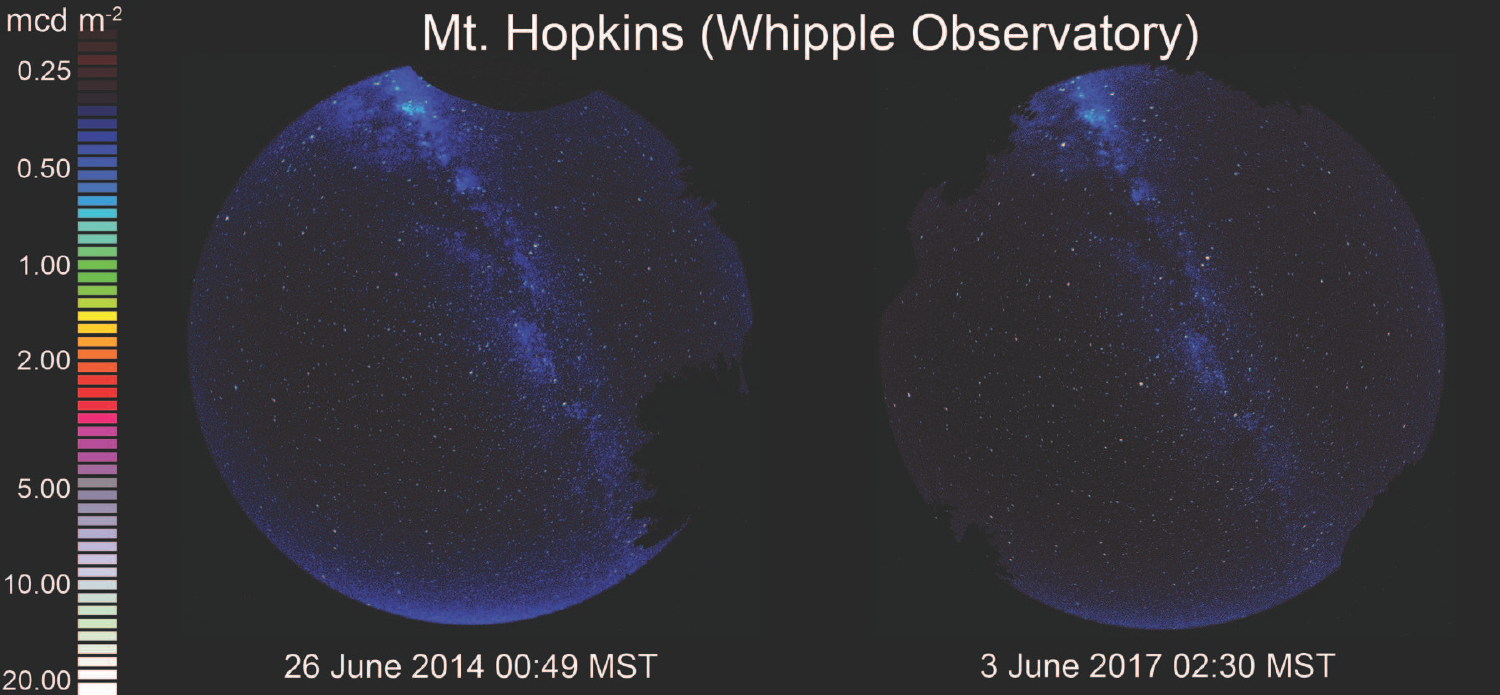}
\caption{(continued)}
\end{figure}
\begin{figure}[H]
\centering
\ContinuedFloat 
\end{figure}

Only three sites in our sample yielded a complete set of measurements in both 2014 and 2017 that meet all of our quality criteria (astronomical darkness; no interference from clouds, dust, or ground light sources; no Milky Way in the zenith): Gates Pass, John F.~Kennedy Park, and the Mount Lemmon SkyCenter. Percent changes in zenith luminance obtained through both direct SQM-L measurements and photometry of all-sky calibrated imagery for these three sites between 2014 and 2017 are given in Table 4.
\begin{table}[tbh]
\medskip
\resizebox{\linewidth}{!}{%
\tabcolsep=2pt
\begin{tabular}{lccc}
 \hline
 \textbf{Site name} & \textbf{Distance (km)} & \textbf{$\Delta$SQM-L (\%)} & \textbf{$\Delta$sSQM-L (\%)}\\
 \hline
 John F.~Kennedy Park & 6.1 & +21.1$^{+4.5}_{-4.4}$ & +0.9$^{+1.9}_{-1.8}$ \\
 Gates Pass & 12.2 & +13.2$^{+9.8}_{-9.0}$ & $-19.1\pm1.5$\\
 Mount Lemmon SkyCenter & 29.8 & $-19.9^{+6.1}_{-5.7}$ & $-17.6\pm1.5$\\
 \hline
 \end{tabular}}
\caption{ Percent changes in zenith illuminances between 2014 and 2017 for three sites (1) for which a complete set of measurements in both exists and that (2) meet all of our quality criteria. ``Distance'' means the radial distance from downtown Tucson (32$^{\circ}$13$^{\prime}$19.3$^{\prime\prime}$N 110$^{\circ}$58$^{\prime}$07.0$^{\prime\prime}$W). To differentiate measurements made with actual SQM-L units from predicted values obtained from synthetic aperture photometry of all-sky imagery, we denote the latter here as ``sSQM-L''.}
\label{junk}
\end{table}

Of these, only one site gives consistent results between SQM-L measurements and predicted SQM-L values obtained from aperture photometry of luminance-calibrated all-sky imagery in the sense that both sources show changes of approximately the same magnitude and sign: the Mount Lemmon SkyCenter, which showed zenith luminance changes between the two epochs of ($-19.9^{+6.1}_{-5.7}$)\% and ($-17.6\pm1.5$)\%, respectively. In all visual skyglow indicators we considered, it appears that the zenith luminance at the Mount Lemmon SkyCenter decreased in the time period that included the bulk of the changes to Tucson's LED lighting. Other measurement sites show either indications of increased brightness or evidence of systematic errors in the 2014 data collection; we suspect that the former were influenced by changes to highly local lighting sources between the two observation epochs.

We hesitate to draw strong conclusions about the impact of the City of Tucson's LED conversion on the brightness of the Tucson light dome, other than to say that we find no evidence that the LED conversion resulted in consistently brighter night skies. The impact to skyglow is either neutral, or possibly toward lower intensities, and depends on location. Further, it is possible that in some cases where apparent large positive changes in zenith luminance exist, skyglow from local ($\sim$few km distant) light sources dominate over the background signal attributable to the skyglow from the integrated light of Tucson and its surroundings.

\begin{figure}[tbh]
\centerline{\includegraphics[width=1.0\linewidth]{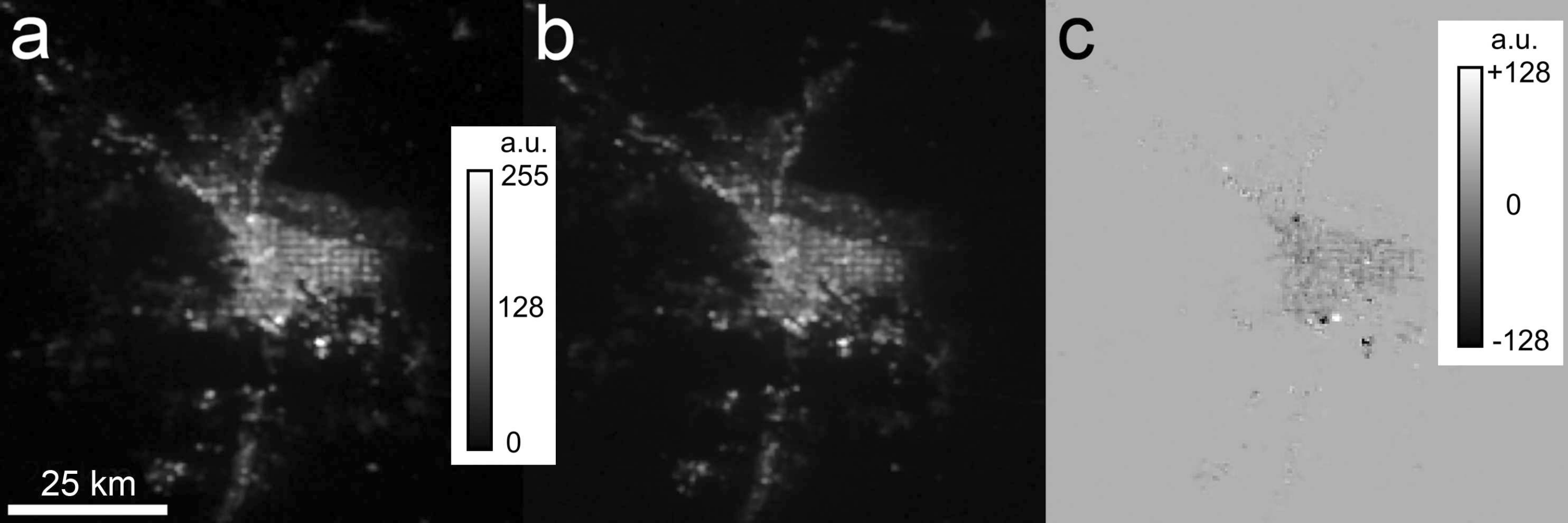}}
\caption{Panels (a) and (b) show VIIRS-DNB monthly cloud-free composite images of Tucson and vicinity in June 2014 and June 2017, respectively. The original 32-bit images have been resampled to 8 bits and scaled identically to allow resolution of individual bright light sources without saturation; scale bars indicating the gray levels in arbitrary units (`a.u.') are provided. The horizontal line in panel (a) is a 25-km scale indicator common to all three panels; the spatial resolution of each image is approximately 750 m pixel$^{-1}$. Panel (c) results from the subtraction of (a) from (b), showing the change in pixel intensity between the two epochs. North is up and east is right in all images.}
\label{dnbcomparison}
\end{figure}

As a point of comparison illustrating the need for ground-based validation of skyglow changes after LED retrofits, we obtained nighttime orbital imagery of Tucson from the VIIRS archive during the two epochs of our ground observations.~Figure~\ref{dnbcomparison} shows VIIRS-DNB monthly cloud-free composite images of Tucson and its surroundings in June 2014 and June 2017, along with a difference image made from the two. The difference image shows an overall apparent decrease in the amount of light emitted by the city core during those three years. Suburban areas serve as controls, since they did not undergo large-scale changes to their lighting during the same period. They show mostly an apparent net zero change, while pixels defining the area of the City of Tucson in which the LED conversion was carried out show a decrease in intensity. 

\begin{figure}[tbh]
\centerline{\includegraphics[width=0.9\linewidth]{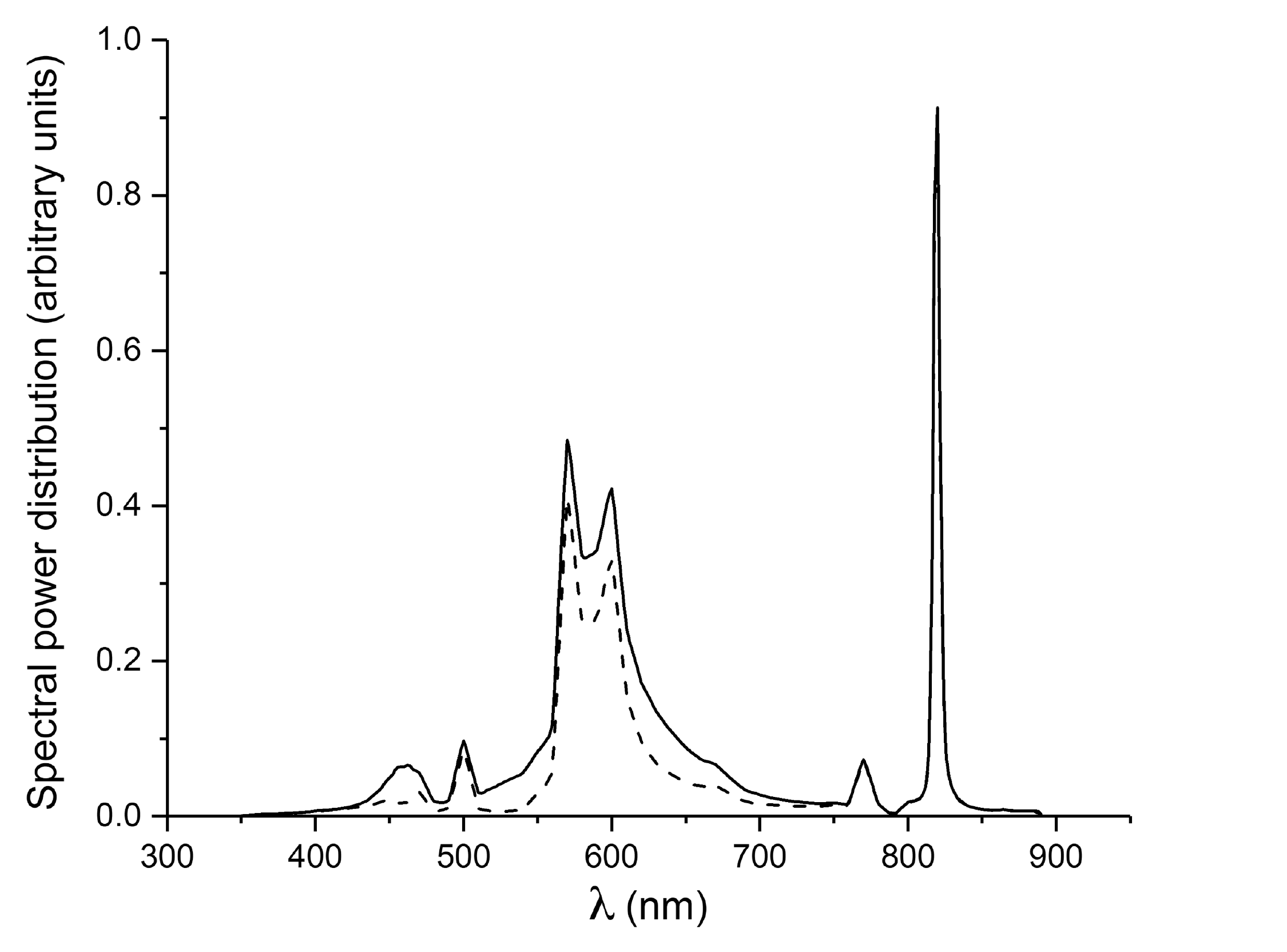}}
\caption{Normalized spectral power distributions for the municipal light emissions of Tucson in the pre-retrofit (dashed line) and post-retrofit (solid line) conditions described in the main text.}
\label{spd}
\end{figure}

To determine numerically the change in light emissions from the part of Tucson receiving LED retrofits between the two epochs, we calculated for each epoch the irradiance ($E$) over the scotopic visual passband and the VIIRS-DNB passband. These irradiances are the integrals of the source spectral power distribution, written here as $P(\lambda)$, in the DNB multiplied by, alternately, the scotopic vision response curve, $S(\lambda)$ \citep{CIE1951}, and the normalized VIIRS-DNB spectral response function, $V(\lambda)$ \citep{VIIRSSRF}:
\begin{align} 
E_S = \int P(\lambda)S(\lambda) d\lambda\\
E_V = \int P(\lambda)V(\lambda) d\lambda,
\end{align}
where each integral is evaluated over the wavelength range 350 nm $\leq$ $\lambda$ $\leq$ 1000 nm. We computed $P(\lambda)$ for the pre- and post-retrofit conditions using the known mixture of sources and their respective fractions of the total municipal light emissions in each case; the assumed spectral power distributions are shown in Figure~\ref{spd}.

The ratio of the irradiances determines the rate by which the DNB data underestimate the signal if part of the light emissions shift toward the blue part of the spectrum, as expected during the conversion to white LED street lighting. The DNB underestimates light emissions from white LED street lights by a factor of about 1.16. Since VIIRS detects light from all sources, we determined this factor under the assumption that direct light from luminaires represents approximately 60\% of all upward-directed radiance. We note further that while there are some near-infrared emission lines in the spectra of high-pressure sodium light sources and that the VIIRS-DNB is sensitive to light from lines at these wavelengths, they contribute less spectral power than light from the 450 nm ``blue peak'' of white LED emissions. 

We find that the upward-directed radiance in the VIIRS-DNB passband originating in the central part of the Tucson metropolitan area apparently changed by -20\% during the LED conversion. However, multiplying the difference by our scaling factor, accounting for the relative insensitivity of the VIIRS-DNB at certain wavelengths, the real change is only about -7\%: ${\Delta}L \propto 1 - (0.8\times1.16) = -0.072$. This figure compares favorably with the -10\% change our models predict for the case of zero uplight ($F=0$\%).

Given that we estimated that the legacy street lighting system comprised 56\% of the total Tucson lumen budget prior to the LED retrofit and that the output of the retrofitted luminaires is 32\% of the earlier system, one might expect a skyglow decrease on the order of 18\% (= 0.56 $\times$ 0.32). That would be correct if all emissions were due only to streetlights. However, the contribution of other sources of non-inventoried light sources, such as buildings and advertising signs, was completely unknown to us. Since these lights are not under direct control of the municipality, we cannot reasonably expect that such lights would be dimmed synchronously with street lighting. Assuming the other lights represent approximately 50\% of all city lights, the skyglow reduction by 7\% (and not 18\%) is realistic. We have simply extracted the maximum information content from an incomplete data set, while keeping in place the theoretical foundations of the procedures used.

We expect that the light-emitting areas of Tucson will continue to expand from year to year according to population growth, implying that an increasing number of light sources will compensate for the decrease predicted theoretically here. As this growth is not strictly predictable, we did not include it in our analysis. The most important outcome from our numerical modeling is that skyglow decreases, independent of the size of the rate of decrease, and that conversion of municipal lighting systems from earlier technologies to LED does not necessarily cause skyglow to increase.

%
%
\section{Summary and Conclusions}\label{sec:summary}

We obtained direct and indirect measurements of the luminance of the night sky from 15 locations in and near Tucson, Arizona, during two epochs in June 2014 and May-June 2017, in between which the City of Tucson converted $\sim$18,000 municipally-owned street lights from a mixture of HPS and LPS to 3000 K white LED, while reducing the number of photopic lumens emitted by the street lighting system by 63\%. We modeled the expected changes in skyglow using the SkyGlow Simulator software, which predicted relative decreases in the brightness of skyglow over Tucson of roughly -10\% for $F=0$\% (full shielding of all lights) and -20\% for $F=5$\% (allowing a small amount of direct uplight). 

The signal corresponding to the change resulting from the street lighting conversion is not entirely clear in our data, but there is some evidence of a decrease at certain measurement sites of the order predicted by the models. Only one site of 15 yielded data of sufficient quality to conclude that the LED conversion reduced skyglow in a manner consistent with our expectations: ($-19.9^{+6.1}_{-5.7}$)\% and ($-17.6\pm1.5$)\%, using two different estimation methods. Upward-directed radiance from the city detected in the VIIRS-DNB shows an apparent decline of ~7\% during the same period. To the extent that upward-directed radiance is a proxy for skyglow intensity observed from the ground, the VIIRS-DNB data may further support the conclusion that skyglow over Tucson decreased after the municipal LED conversion. Therefore, there is some evidence that the lumen reduction accompanying Tucson's conversion to SSL measurably decreased both the intensity of skyglow and upward-directed radiance.

\subsection{Limitations of this work}

There are a number of shortcomings in the approach to this study that could be overcome in future efforts to characterize the outcomes of LED conversions with regard to skyglow over cities. These deficiencies result from the serendipitous nature of the 2014 field observations, which were not carried out with the intent of making comparisons to a later epoch. We used different SQM devices in the two campaigns, although we attempted to understand any systematic differences in their responses after the fact through mutual inter-comparison under controlled conditions. A number of the 2014 measurements were carried out under circumstances that were not ideal for the goals of this study; in particular, some of the measurements were made at times during which we now know the data were influenced by either moonlight or twilight interference, or were taken in the presence of clouds. Conditions were not precisely the same during the two epochs, such as the time of night during which measurements were obtained, ambient air temperature, relative humidity, and atmospheric turbidity. Furthermore, not all observations could be precisely replicated in terms of geographic location, and we cannot rule out the possibility that observations from a given site in one epoch or the other were contaminated by the presence of nearby, highly local outdoor lighting sources. Lastly, observations in 2014 and 2017 were undertaken by different observers with different levels of experience, which may have further influenced certain measurements, in particular the naked-eye limiting magnitude estimates.

From these considerations, we conclude that such experiments are difficult to conduct under real-world conditions when carried out in campaign-style fashion. The variability of conditions from one night to another, not to mention nights separated by several years, is sufficiently unpredictable that perfect comparisons are not possible. Rather, and in hindsight, a more effective approach would be to install permanently-mounted sky brightness monitors prior to the start of an LED conversion, and to collect data throughout. Not only would such an approach allow for the suppression of variable conditions from one night to the next, but it would also offer temporal resolution that could be related back to the schedule of luminaire replacements in a city.

The number and distribution of monitors should also be sufficient to adequately sample the spatial distribution of light emitted or reflected into the night sky over the geographic extent of a city of arbitrary size. The recent work of Bar\'{a} \citep{Bara2017} suggests that optimum spatial sampling to reconstruct the zenith sky luminance to a precision of $\sim$10\% rms is about one sample per square kilometer.

\subsection{Applicability to other cities}

Tucson is unusual compared to many world cities due to its typically low relative humidity (and therefore often low AOD). On average, its frequently transparent night skies lead to darker conditions than experienced by other cities in more humid climates. However, it can at times be a dusty environment. We accounted for turbidity effects in our models that can be likewise extended to models of other cities, given local AOD inputs.

In policy terms, Tucson is also unusual in that the concern for limiting light pollution is connected to the site protection of astronomical observatories that contribute significantly to its local economy. Most world cities do not have such an influence on local decision making with respect to outdoor lighting practices and policies. Yet the approach taken by Tucson of dimming its LED streetlights relative to the light levels of its legacy HPS/LPS system could be effectively implemented by any city interested in limiting increases in skyglow during a conversion to SSL. There are preliminary indications in this study that that reducing lumens during municipal LED conversions may reduce skyglow over a city, even given a shift in the spectrum of lighting toward bluer wavelengths. Presumably, full shielding of luminaires is especially effective in this regard; we expect that cities moving from partially shielded luminaires to fully shielded ones during their LED conversions will see even greater overall decreases in skyglow when coupled with Tucson-like dimming schemes. However, cities that elect not to dim may still see some benefit by simply converting to modern, fully shielded luminaires. 

\subsection{Future work}

There is a distinct need for further studies like this one, given that policymakers planning LED conversions for their jurisdictions are confronted with decisions that affect outcomes for skyglow, whether positive or negative. We encourage more before/after studies, especially in cases where cities either (1) keep the intensity of their new lighting systems equal to the intensity of their previous systems, or (2) increase or decrease intensity during the conversion. We expect skyglow will worsen in cases where the `rebound' effect results in the installation of more lighting than existed prior to conversion. The installation of permanent sky brightness monitoring equipment can obviate some of the practical problems identified in our single-epoch observations by allowing researchers to more readily identify typical or average conditions, as well as to understand the range of parameters such as AOD. 

We also encourage monitoring of upward-directed radiance in VIIRS-DNB before and after LED conversions as a check on the ground-based skyglow measurements. It is possible to use the approach of Falchi et al.~\citep{Falchi2016} in predicting skyglow changes using DNB radiances as inputs and a model for skyglow formation, while keeping in mind the relative insensitivity of the DNB to emissions in the 450 nm ``blue spike'' of white LED products. 

Lastly, it would be interesting to examine Tucson road safety and uniform crime statistics after 2017-18 data become available to see whether changing the illuminance of roadways had any effect on either traffic accidents or the perpetration of certain crimes. However, given the criticism of poorly-constructed studies aiming to find correlations between crime, public safety, and outdoor lighting \citep{Marchant2004,Marchant2005,Marchant2017}, such efforts should be very carefully considered.

\section*{Acknowledgments}

\noindent
The authors wish to recognize Jesse Sanders (LED Project Manager, City of Tucson, Arizona) for providing streetlight inventory data, and Gilbert Esquerdo (Fred Lawrence Whipple Observatory and Planetary Science Institute) for assistance in obtaining some sky luminance data during the 2017 measurement campaign. They further thank the two anonymous reviewers whose comments and suggestions improved the quality of this manuscript. \\

\noindent
Funding: This work was supported by the Slovak Research and Development Agency under contract No. APVV-14-0017.  Computational work was supported by the Slovak National Grant Agency VEGA (grant No. 2/0016/16). 

\bibliography{barentine-et-al}

\end{document}